\documentstyle[preprint,eqsecnum,aps]{revtex}

\begin{document}

\preprint{HD-TVP-97/10}

\title{\vspace{1.5cm} $\pi$-$K$  scattering lengths at
      finite temperature in the Nambu--Jona-Lasinio model}

\author{P.~Piwnicki,  S.\,P.~Klevansky and P.~Rehberg }
\address{ Institut f\"ur Theoretische Physik, \\
          Philosophenweg 19, D-69120 Heidelberg, Germany}

\maketitle
\vspace{1cm}

\begin{abstract}
The transition amplitude for $\pi K$ scattering is evaluated within the
SU(3) Nambu--Jona-Lasinio model. Ordering terms according to the
expansion in $1/N_c$ leads to a box-like diagram, $t$ channel diagrams
that admit scalar isoscalar $(\sigma,\sigma')$ exchanges, and a $u$
channel exchange of a scalar isodoublet $\sigma_K$ that has quantum
numbers corresponding to the $K_0^*(1430)$. Both the Pauli-Villars and
O(3) regularization procedures are used to evaluate the $T=0$ values of
the $l=0$ scattering lengths $a_0^{3/2}$ and $a_0^{1/2}$. The finite
temperature dependence is studied. We find that the variation in the
$t$ channel  in the calculation of $a_0^{3/2}$ 
  leads to a change
in $a_0^{3/2}$ of a factor of about two over the temperature range of
$T=150$~MeV.
\end{abstract}
\clearpage

\section{Introduction}
The pseudoscalar octet $(\pi,K,\eta)$ forms the set of lowest mass 
hadronic states and are thus particles that play a central role in heavy ion
collisions, since they  occur copiously.
 In particular the pion, with lowest mass and being an isospin triplet
has a central role, while the kaonic sector, occurring as isospin doublets,
gives us information  on strangeness production.    Since this octet
 consists of the
 Goldstone bosons associated with spontaneous chiral  symmetry
breaking,  their role in nature is determined almost exclusively by this
feature.   Conversely, our understanding of chiral symmetry breaking is 
augmented by studying the properties of such mesons, and this can be
extended to finite temperatures and baryon density.

The purpose of this paper is to examine elastic $\pi K\rightarrow \pi K$
scattering in particular at $T=0$ and at finite temperatures.
Expressions for the transition amplitudes and hence the 
 scattering lengths at $T=0$ are derived and then
the temperature dependence of the scattering lengths is investigated.
This is of interest (a) in its own right, although at present it is 
unclear how one would access finite temperature scattering lengths
experimentally, and (b) as it provides a first step in the calculation
of the elastic $\pi K\rightarrow \pi K$ cross-sections as a function of the
temperature that are required as input for the collision dynamics for a 
simulation of heavy-ion collisions using a chiral Lagrangian.  

At $T=0$, $\pi K$ scattering has been studied within the framework of
chiral perturbation theory (CHPT) \cite{bkm1,bkm2,bkm3}.   
    It is not {\it a priori} 
clear to what extent a chiral expansion is appropriate for properties
involving the strange quark, in view of its mass.   Yet the results of
even the early calculations of the scattering lengths\footnote{
There is an early Russian publication on this subject that we also draw
the reader's attention to \cite{vo}} \cite{bkm1}
 fall within the
expected measured range \cite{mat,joh,kar},
 although one should note that actual experimental
values are rather poorly  known.  An elaboration of this approach that 
includes intermediate resonances \cite{bkm2},
 in particular the exchange of the  vector
$K^*$ in the $u$-channel as well as of a $\rho$-meson in the $t$-channel,
alters the actual numerical values by 2\% at most.

In this paper, we wish to examine the $\pi K$ scattering lengths at
finite temperature.   The Nambu--Jona-Lasinio (NJL) model \cite{reviews,hatkun}
 gives a good
description of these deeply bound mesons at $T=0$.   In obtaining such
a description, the mean field or Hartree approximation to the self-energy,
together with the random phase approximation for the mesonic fields has been
used.   It has been recognized that these approximations taken together 
constitute an expansion in the inverse number of colors $1/N_c$, within the
model \cite{quack,dmitra} 
and are essential in order to preserve the chiral symmetry of the
underlying Lagrangian.
For $T\ne0$, the 
dependence of some of the pseudoscalar and scalar mesons as a function of
temperature has qualitatively been validated by lattice gauge calculations 
\cite{laer1,laer2}.
The results of Ref.\cite{laer2} display a remarkable qualitative agreement
with the finite temperature behavior of the NJL model (pole) masses.
This gives one some confidence in the application of this model in this
form to finite temperature, although it is speculative.
 
In addition, it can be shown that the current algebra results or
alternatively the results of chiral perturbation theory to lowest order can
be obtained from this model on making a suitable chiral expansion
\cite{muller,lemmer}.
   It is then a simple issue to
extend calculations to finite temperature, and examine the consequences
thereof. 
There is no necessity
 to enforce the chiral limit, since this is a model study.    
  We thus examine the three flavor NJL
model.      Contributions to the scattering amplitude are organized
according to the  expansion in  $1/N_c$.
    To lowest order 
within this scheme, there are several sorts of graphs that
occur: a box graph is present, analogous to the four-point tree-like
structure of mesonic theories.   In addition, in the $t$ channel,
exchange of neutral scalars, in this case the $\sigma_0$ and $\sigma_8$
$\rightarrow \sigma$ and $\sigma'$ states occur.    This has no
counterpart in the CHPT version \footnote{Note that no $\rho$-meson can
be exchanged in this minimal version of the NJL model, but would also
be of the same order as the $\sigma$ exchanges 
in  the $1/N_c$ expansion within an extended
NJL model \cite{lemmer}.
   For this reason, we do not calculate the $l=1$
scattering lengths $a_1^{1/2}$, $a_1^{3/2}$.}. Finally, the $u$ channel
contains the exchange of a $\sigma_K$, corresponding to the $K_0^*$ meson --
note that this also differs from the CHPT calculations that explicitly
 include vector
resonance exchange of the $K^*$  \cite{bkm2}.
Note that in our case, no six point tadpole-like  diagrams \cite{bkm1} 
are present.    In addition,
unitarising
graphs that include $\pi K$ scattering in the intermediate state 
are not 
included to this lowest order in $1/N_c$ calculation of the scattering
lengths in the NJL model, but which would be essential for a complete
description of the scattering amplitude of
 cross section calculated as a function of center of mass energy.

Before discussing our results, we comment on the procedure that is to be
taken, in particular with regard to the application of the mean field plus
random phase approximation at $T=0$ and at $T\ne0$.   As discussed earlier,
this coupled symmetry preserving approximation works excellently at $T=0$. 
This is true for calculating static mesonic properties, such as masses, mean
charge radii and scattering lengths, particularly in the SU(2) sector, where
these have been studied in detail and are well-known.    Corrections to the
mean field values introduce chiral logarithms,
see for example \cite{hippe} and further powers in the pion mass.
Calculations of the full scattering cross-sections, however, do require
these corrections, even at $T=0$, in order to preserve unitarity.   Thus
only scattering lengths can be reliably calculated.    This is
well-established for the two flavor case \cite{bandco} and is the
reason for our restriction to investigating this quantity only.

   In extending our calculation to finite temperature, we are being
speculative, encouraged however, by the success of the 
qualitative agreement of the lattice predictions
of the temperature variation of particle masses with those predicted by the 
model.    Furthermore, it is not expected that terms beyond
this leading $1/N_c$ calculation play a dominant role, since the chiral 
logarithms and powerlike mass terms that might then occur
depend then on the temperature dependent pion mass, which in turn stays
constant within the model for a wide range of temperatures, even up to
200 MeV.   The divergences that have been  found in this work in the scattering
lengths are induced by the presence of the inverse of the decay constants
which themselves become small at high temperatures.   This feature should 
remain unchanged by higher order corrections.

Our model results depend on the regularization procedure used, and can
therefore at most be qualitative.     While the Pauli-Villars method is
generally preferred at $T=0$ since it preserves all symmetries,  and
enables one to check that the crossing symmetries are correctly
implemented, at $T\ne 0$, Lorentz invariance is broken by the heat bath
so that a three dimensional cutoff can be used.
We thus only expect to obtain a  qualitative
description.    We evaluate the scattering
lengths at $T=0$ using both prescriptions, and considering parameters
sets of other authors \cite{bv,pr,hatkun} that were
 used for different purposes to fix other
variables.    In this sense, we do not fit model parameters so as to
explicitly give us good values of the scattering lengths.    Our
results at $T=0$ are obtained using both regularization schemes.
  At finite temperatures, we find that the $t$ channel
varies somewhat with temperature due to the exchange of the scalar
mesons, which themselves have a strong temperature dependence, and we
find that the variation of $a_0^{3/2}$ in particular, is visible.
 At $T=150$~MeV,  the $T=0$ value is increased  by a
factor of 2.36.    This is in contrast to the temperature dependence found
for the $\pi\pi$ scattering lengths \cite{qufamily} which show almost no
change over $T=0$ to 150 MeV.
 The value of $a_0^{1/2}$ changes only slightly, increasing in this case
only by a factor of 1.18.

This paper is organized as follows.   In Section~\ref{sec2}, we give a
brief introduction to the NJL Lagrangian, sufficient to introduce our
notation.  In Section~\ref{sec3}, we classify the scattering diagrams
required for elastic $\pi K$ scattering within the framework of the
model and examine in particular the reaction $\pi^+  K^+\rightarrow
\pi^+  K^+$.    This is studied explicitly at $T=0$ in Sec.~\ref{sec4a}
and the scattering
lengths for this process as well as the remaining $\pi K$ processes,
which can be obtained via crossing symmetry, are then evaluated.  In
Sec.~\ref{sec4}, the generalization to finite temperature is
discussed.    We summarize and conclude in Sec.~\ref{sec5}.

\vskip 0.5in

\section{SU(3) Nambu--Jona-Lasinio model} \label{sec2}
The SU(3) NJL model is well documented elsewhere \cite{reviews,hatkun}. 
 In this section,
we give only a brief description of the  model in order to introduce
our notation. Our starting point is the minimal three flavor chiral Lagrangian
density
\begin{eqnarray}
{\cal L}_{{\rm chiral}} = \bar\psi i\!\not\! \partial \psi &+& G \sum_{a=0}^8
\left[(\bar\psi \lambda^a\psi)^2 + (\bar\psi i\gamma_5 \lambda^a\psi)^2
\right] 
\nonumber \\
 &-& K\left[\rm{det}(\bar\psi(1+\gamma_5)\psi) + \rm{det}(\bar\psi (1-\gamma_
5)\psi)\right] 
\label{e:lag}
\end{eqnarray}
that contains a four point coupling with strength $G$, and six point
coupling, with coupling strength $K$.    Color, flavor and spinor
indices of the quark fields $\psi$ are suppressed.   The term moderated
by $G$ contains the Gell-Mann matrices $\lambda^a$, $a=1..8$ plus the
matrix $\lambda^0=\sqrt{2/3}I$, where $I$ is the unit matrix, thereby
ensuring the U(3) symmetry of this term.   This symmetry is broken by
the determinantal term, motivated by instanton effects, down to the
chiral group $SU_L(3)\times SU_R(3)$.   In order to incorporate the
known current quark masses, an additional term that explicitly breaks
the chiral symmetry,
\begin{equation}
{\cal L}_{{\rm mass}} =-( m_u^0 \bar u u + m_d^0 \bar d d + m_s^0\bar s s)
\label{e:mass}
\end{equation}
must be introduced, so that ${\cal L}_{{\rm NJL}} = {\cal L}_{{\rm chiral}}
+ {\cal L}_{\rm {mass}}$.

A straightforward evaluation of the self-energy in the mean field
approximation yields constituent quark masses that are modified from their
current values due to dynamical symmetry breaking \cite{bv,reviews}
\begin{equation}
m_i = m_i^0 + 4iGN_c {{\rm tr}} S^i - 2KN_c^2({\rm tr} S^j)({\rm tr}
 S^k), \quad\quad i\ne j\ne k \ne i \quad ,
\label{e:self}
\end{equation}
where the propagator $S^i(x,x')$ satisfies
\begin{equation}
(i\!\not\!\partial_x - m_i) S^i(x,x') = \delta^{(4)}(x-x')
\label{e:eq}
\end{equation}
in the mean field approximation and $i$ is a flavor index.

The mesonic sector is viewed most simply from the effective NJL Lagrangian
that is obtained on constructing an effective four point interaction from
the six point term, on contraction of two fermionic fields.   One has
\begin{eqnarray}
{\cal L}_{{\rm chiral}} = \bar \psi i\!\not\!
\partial \psi &+& \sum_{i=0}^8
\left[K_i^{(-)} (\bar \psi\lambda^i\psi)^2 + K_i^{(+)}(\bar\psi i\gamma_5
\lambda^i\psi)^2\right] \nonumber \\
&+&\left[\frac 12 K_m^{(-)}(\bar\psi\lambda^8\psi)(\bar\psi\lambda^0\psi)
+\frac 12 K_m^{(+)}(\bar\psi i\gamma_5\lambda^8\psi)(\bar\psi 
i\gamma_5\lambda^0\psi)\right] \nonumber \\
&+&\left[\frac 12 K_m^{(-)}(\bar\psi\lambda^0\psi)(\bar\psi\lambda^8\psi)
+\frac 12 K_m^{(+)}(\bar\psi i\gamma_5\lambda^0\psi)(\bar\psi 
i\gamma_5\lambda^8\psi)\right] \nonumber \\
\label{e:eff}
\end{eqnarray}
with the effective couplings
\begin{eqnarray}
K_0^{(\pm)} &=& G\mp \frac 13 N_c K(i {\rm tr} S^s + 2i{\rm tr}
 S^u) \nonumber \\
K_1^{(\pm)} = K_2^{(\pm)} = K_3^{(\pm)}&=& G\pm \frac 12 N_c Ki {\rm tr} S^s 
\nonumber \\
K_4^{(\pm)} = K_5^{(\pm)} =K_6^{(\pm)} = K_7^{(\pm)} &=& G \pm \frac 12
N_c Ki {\rm tr} S^u \nonumber \\
K_8^{(\pm)} &=& G \mp \frac 16N_c K(i {\rm tr} S^s - 4 i {\rm tr}
 S^u) \nonumber \\
K_m^{(\pm)} &=&  \mp \frac{\sqrt{2}}{3} N_c K (i{\rm tr} S^s - i {\rm tr} S^u)
\quad , \label{e:couplings}
\end{eqnarray}
in the case $m_u^0 = m_d^0$.     Both scalar and pseudoscalar octet 
channels may be identified by the isospin operator correspondence
\begin{eqnarray}
\lambda_3 &\rightarrow& \pi^0, \sigma_\pi \nonumber \\
\frac 1{\sqrt 2}(\lambda_1 \pm i\lambda_2) &\rightarrow& \pi^{(\pm)},
\sigma_\pi^{(\pm)} \nonumber \\
\frac 1{\sqrt 2}(\lambda_6 \pm i\lambda_7) &\rightarrow& K^0, \bar K^0,
\sigma_K^0, \bar \sigma_K^0 \nonumber \\
\frac 1{\sqrt 2} (\lambda_4 \pm i\lambda_5) &\rightarrow& K^{(\pm)},
\sigma_K^{(\pm)} \quad ,
\end{eqnarray}
using an obvious notation for the scalar sector.  Experimentally,
one could tentatively assign the scalar particles  $(\sigma_\pi,
\sigma_K)\rightarrow (a_0(980), K_0^*(1430))$.
The assignment of the $(\sigma,\sigma\prime)$ to actual particles such as
the controversial $f_0(400-1200)$ or $\sigma$ \cite{pdg} or $f_0(980)$
and $f_0(1300)$ is uncertain and we thus 
     retain our model notation
here.   The pionic and
kaonic scalar and pseudoscalar sectors can be treated independently now.
The quark-antiquark
scattering amplitude, evaluated in the random phase approximation is
\begin{equation}
 M_{\pi,\sigma_\pi} = \frac{2K_1^{(\pm)}}{1-2K_1^{(\pm)}\Pi_{P,S}^{q\bar q}
(k^2)}
\label{e:m}
\end{equation}
for the pions, or
\begin{equation}
 M_{K,\sigma_K} = \frac{2K_4^{(\pm)}}{1-2K_4^{(\pm)}\Pi_{P,S}^{s\bar q}
(k^2)} \quad ,
\label{e:mk}
\end{equation}
for the kaons, expressed in terms of the irreducible polarization
\begin{equation}
-i\Pi_{P,S}^{f_1f_2}(k^2) = - 2N_c \int \frac{d^4q}{(2\pi)^4} {\rm tr} [i
S^{f_1}(p+q) \Gamma^{P,S} iS^{f_2}(q) \Gamma^{P,S}] \quad .
\label{e:pol}
\end{equation}
Here $f_1$ and $f_2$ refer to the quark flavors  $q=u,d$ and $s$.
   The indices $P,S$ refer to
pseudoscalar or scalar functions respectively, and $\Gamma^P=i\gamma_5$,
while $\Gamma^S=1$.   The meson masses with the appropriate quantum 
numbers are obtained by solving
\begin{eqnarray}
1-2K_1^{(\pm)} \Pi_{P,S}^{q\bar q}(k^2) &=& 0 \nonumber \\
1-2K_4^{(\pm)} \Pi_{P,S}^{s\bar q}(k^2) &=& 0
\label{e:mmass}
\end{eqnarray}
for the pionic or kaonic masses respectively, 
while the associated couplings are given as
\begin{equation}
g_{M_\pi qq}^2 = \left(\frac{\partial \Pi^{q\bar q}_{P,S}}{\partial k^2}\right)
^{-1}_{k^2=m_{M_\pi}^2} \quad \quad
g_{M_Ksq}^2 = \left(\frac{\partial \Pi^{s\bar q}_{P,S}}{\partial k^2}\right)
^{-1}_{k^2=m_{M_K}^2} \quad .
\label{e:gs}
\end{equation}
In this equation, $M_\pi$ refers to the mesons $\pi$ or ${\sigma_\pi}$ 
while $M_K$ refers  to the mesons $K$ or $\sigma_K$  as required.
The mesons $\eta$ and $\eta'$, or correspondingly the scalars $\sigma$ and
$\sigma'$ are coupled.  Here the quark-antiquark 
scattering amplitude in either sector
takes the matrix form
\begin{equation}
M=2K(1-2\Pi K)^{-1}
\label{e:matrx}
\end{equation}
where
\begin{eqnarray}
\Pi = \left(\begin{array}{cc}
 \Pi_{00} & \Pi_{08} \\ \Pi_{80} & \Pi_{88} 
\end{array}\right)
\label{e:piarray}
\end{eqnarray}
with
\begin{eqnarray}
\Pi_{00} &=& \frac 13(2\Pi^{qq} + \Pi^{ss}) \nonumber \\
\Pi_{08}=\Pi_{80} &=& \frac{\sqrt2}3(\Pi^{qq} - \Pi^{ss}) \nonumber \\
\Pi_{88} &=& \frac 13(\Pi^{qq} + 2\Pi^{ss})
\label{e:pis}
\end{eqnarray}
and
\begin{eqnarray}
K &=& \left(\begin{array}{cc}
 K_{00} & K_{08} \\ K_{80} & K_{88}
\end{array}\right), \\
  K_{ii} &=& K_i^\pm, 
K_{08} = K_{80} = \frac 12 K_m^\pm
\label{e:karray}
\end{eqnarray}
in either the scalar or pseudoscalar sector.   For the purposes of our 
calculation of $\pi K$ scattering lengths,
 we will not require information on the $\eta$ and $\eta'$
mesons.    However, the exchange of a $\sigma$ or $\sigma'$ in the
intermediate state within $\pi K$ scattering may occur.    We thus 
require the quark-antiquark scattering amplitude for the exchange of the
$\sigma$ and $\sigma'$.   This can be written as
\begin{equation}
D_{\sigma,\sigma'} = M_{00}\lambda_0\times\lambda_0 +
M_{08}\lambda_0\times
\lambda_8 + M_{80}\lambda_8\times\lambda_0 + M_{88}
\lambda_8\times\lambda_8 \quad .
\label{e:dsss}
\end{equation}
Since the neutral $\sigma$ and $\sigma'$ are both exchanged in any
single process, it is not necessary to resolve this scattering amplitude
into the individual components for these particles via
diagonalization.   For our purposes, the individual masses and
couplings are not required, although this provides an alternative
method.   The reader is referred to Refs.~\cite{hatkun,pr},
 who perform such a
diagonalization explicitly.

\section{Scattering lengths for $\pi K\rightarrow \pi K$ at $T=0$}
\label{sec3} In general, if one regards the process $a+b\rightarrow
a'+b'$, the scattering amplitude for a fixed isospin $I$ can be
decomposed into partial wave amplitudes, {\it i.e.}
\begin{equation}
T^I(s,t,u) = 16\pi \sum_{l=0}^\infty (2l+1) t_l^I(s) P_l(\cos \theta)
\label{e:ts}
\end{equation}
where $s$, $t$ and $u$ are the usual Mandelstam variables $s=(p_a+p_b)^2$,
$t = (p_a-p_{a'})^2$ and $u=(p_a-p_b')^2$, and $\theta$ is the scattering
angle.   The phases $\delta_l(s)$ parametrize the scattering amplitude as
\begin{equation}
t_l^I(s) = \frac{\sqrt{s}}{2q}\frac 1{2i} \left[e^{2i\delta_l^I(s)}-1\right]
\quad , \label{e:littlets}
\end{equation}
where
\begin{equation}
q = \frac{1}{2\sqrt{s}}
\sqrt{\left[s-(m_a+m_b)^2\right]\left[s-(m_a-m_b)^2\right]}
\end{equation}
is the center of mass momentum of the incoming particles.
The differential cross section is
\begin{equation}
\frac{d\sigma^I}{d\Omega} = \frac 1{64\pi^2 s}|T^I(s,t)|^2 \quad ,
\label{e:dsigma}
\end{equation}
while the total cross section can be expressed purely in terms of the
phase shifts, 
\begin{equation}
\sigma^I = \frac{4\pi}{q^2}\sum_l(2l+1)\sin^2\delta_l^I(s) \quad .
\label{e:sigma}
\end{equation}
At low energies, the partial wave amplitudes can be expanded in the form
\begin{equation}
\Re\left[ t_l^I(s)\right] =
\frac{\sqrt{s}}2 q^{2l}[a_l^I + b_l^I q^2 + O(q^4)] \quad ,
\label{e:expansion}
\end{equation}
defining the scattering amplitudes $a_l^I$ and the slope parameters $b_l^I$.
At the kinematic threshold $q\rightarrow 0$, the cross sections in each
isospin channel are given purely in terms of the $s-$wave amplitudes,
{\it i.e.}
\begin{equation}
\sigma^I = 4\pi \left(a_0^I\right)^2 \quad .
\label{e:limitsigm}
\end{equation}

Our interest is in  the $\pi K$ system, and we  start 
with the process of maximal
isospin, $I=3/2$,
\begin{equation}
\pi^+ + K^+ \rightarrow \pi^+ + K^+ \quad .
\label{e:max}
\end{equation}
By exchanging particles $a$ and $a'$, one directly obtains the scattering
process 
\begin{equation}
\pi^- + K^+ \rightarrow \pi^- + K^+ \quad ,
\label{e:min}
\end{equation}
in which $u$ is the center of mass variable.   This reaction contains not 
only an $I=3/2$ component, but  the $I=1/2$ component also.   Crossing
matrices relate the various amplitudes.  Using an obvious notation,   
one may write
\begin{eqnarray}
T_s^I(s,t,u) &=& \sum_{I'=1/2}^{3/2} C_{I,I'}^{su} T_u^{I'}(s,t,u)
\nonumber\\
&=&C_{I,1/2}^{su} T_u^{1/2}(s,t,u) + C_{I,3/2}^{su}T_u^{3/2}(s,t,u)
\quad , \label{e:tarray}
\end{eqnarray}
with the crossing matrix \cite{petersen}
\begin{eqnarray}
C_{I,I'}^{su} = \frac13\left(\begin{array}{cc}
        -1 & 4 \\
         2 & 1 \end{array} \right) \quad .
\label{e:cmatrix}
\end{eqnarray}
In practice, this tells us that
\begin{equation}
T^{1/2}(s,t,u) = \frac 32 T^{3/2}(u,t,s) - \frac12 T^{3/2}(s,t,u)
\label{e:relation}
\end{equation}
so that the determination of scattering in the isospin channel $I=3/2$
is sufficient to obtain all information, including the amplitude in the 
 $T=1/2$ channel.
We therefore examine the process $\pi^+  K^+ \rightarrow \pi^+  K^+$
in what follows. 

\subsection{General Classification of Diagrams}
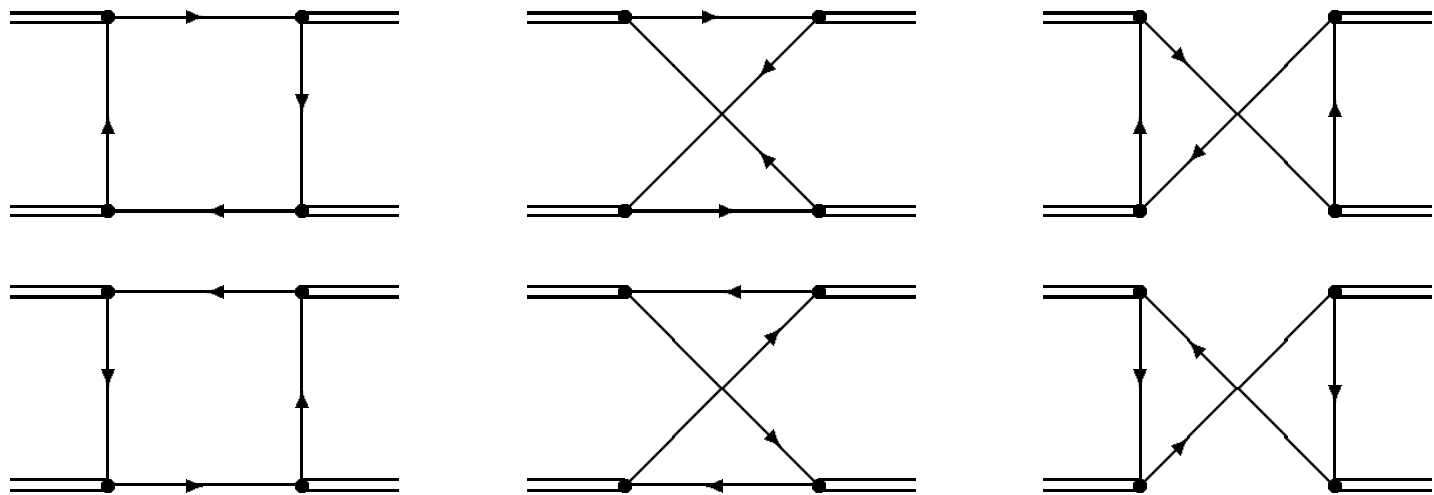
\begin{figure}
\caption[]{Six possible box diagrams that contribute to
           meson-meson scattering.    Double lines denote mesons,
while the internal single lines depict the quark structure.}
\label{fig1}
\end{figure}

Since the coupling strengths in the NJL model, $G\Lambda^2$ and
$K\Lambda^5$ are large, a perturbative expansion in the coupling
strengths is inadmissable.   An alternative expansion in the inverse
number of colors $N_c$ is generally made \cite{quack,dmitra}.
   In this expansion,
fermion loops contribute a factor of $N_c$, while the scattering
amplitudes with mesonic intermediate states contribute a factor of
$1/N_c$.     Thus the diagrams that are leading in the $1/N_c$
expansion, and which are of the same order in this expansion, are both
the box diagrams of the type displayed in Fig.~\ref{fig1} and the
mesonic exchange graphs of Fig.~\ref{fig2}.
 
\subsection{Contributions to $\pi^+  K^+\rightarrow \pi^+  K^+$.}
\subsubsection{Box diagrams}   For the process $\pi^+ K^+\rightarrow \pi^+K^+$,
only one of the box diagrams of Fig.~\ref{fig1}
 that is commensurate with isospin and
strangeness conservation survives.   This is depicted, togther with the
appropriate kinematic variables, in Fig.~\ref{fig3}.    A direct
translation of this diagram leads to the expression
\begin{eqnarray}
iT_{{\rm box}} &=& -4 (ig_{\pi qq})^2 (ig_{Ksq})^2 N_c \label{e:box} \\
&\times & \int\frac{d^4q}{(2\pi)^4}
{\rm tr}\left[i\gamma_5 iS^q(q) i\gamma_5 iS^q(q-p_1)
i\gamma_5 iS^q(q-k_2+k_1) i\gamma_5 iS^s(q-k_2)\right] \quad .
\nonumber
\end{eqnarray}
Here the factor 4 arises from the flavor
algebra.   It has also been assumed that the translation of the Feynman
graph gives $iT$.

\subsubsection{Meson exchange graphs}  
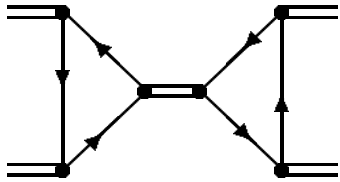
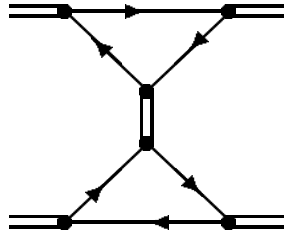
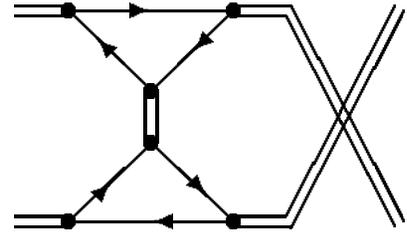
\begin{figure}
\caption[]{Three possible meson exchange graphs,
           corresponding to $s$, $t$ and $u$ channels respectively.
The double lines represent mesons and the single lines represent quarks.}

\label{fig2}
\end{figure}

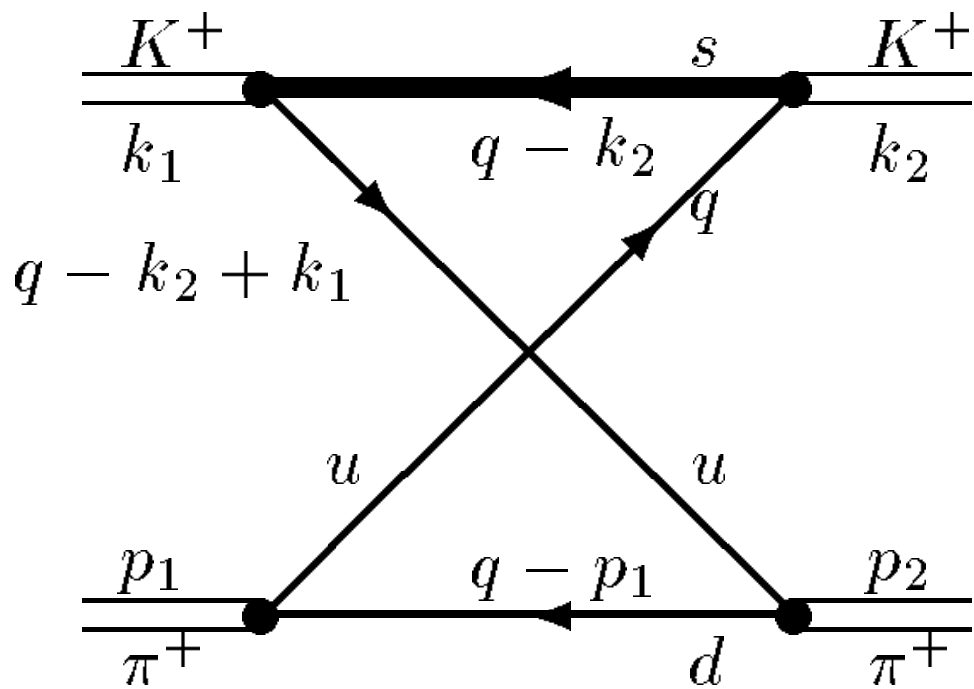
\begin{figure}
\caption[]{The only box diagram  contributing to the elastic scattering
           amplitude $\pi^+K^+\rightarrow \pi^+K^+$.
Here we denote the mesonic states with double lines and quarks by single
lines.   The $s$ quark is denoted by a heavy single quark.}
\label{fig3}
\end{figure}

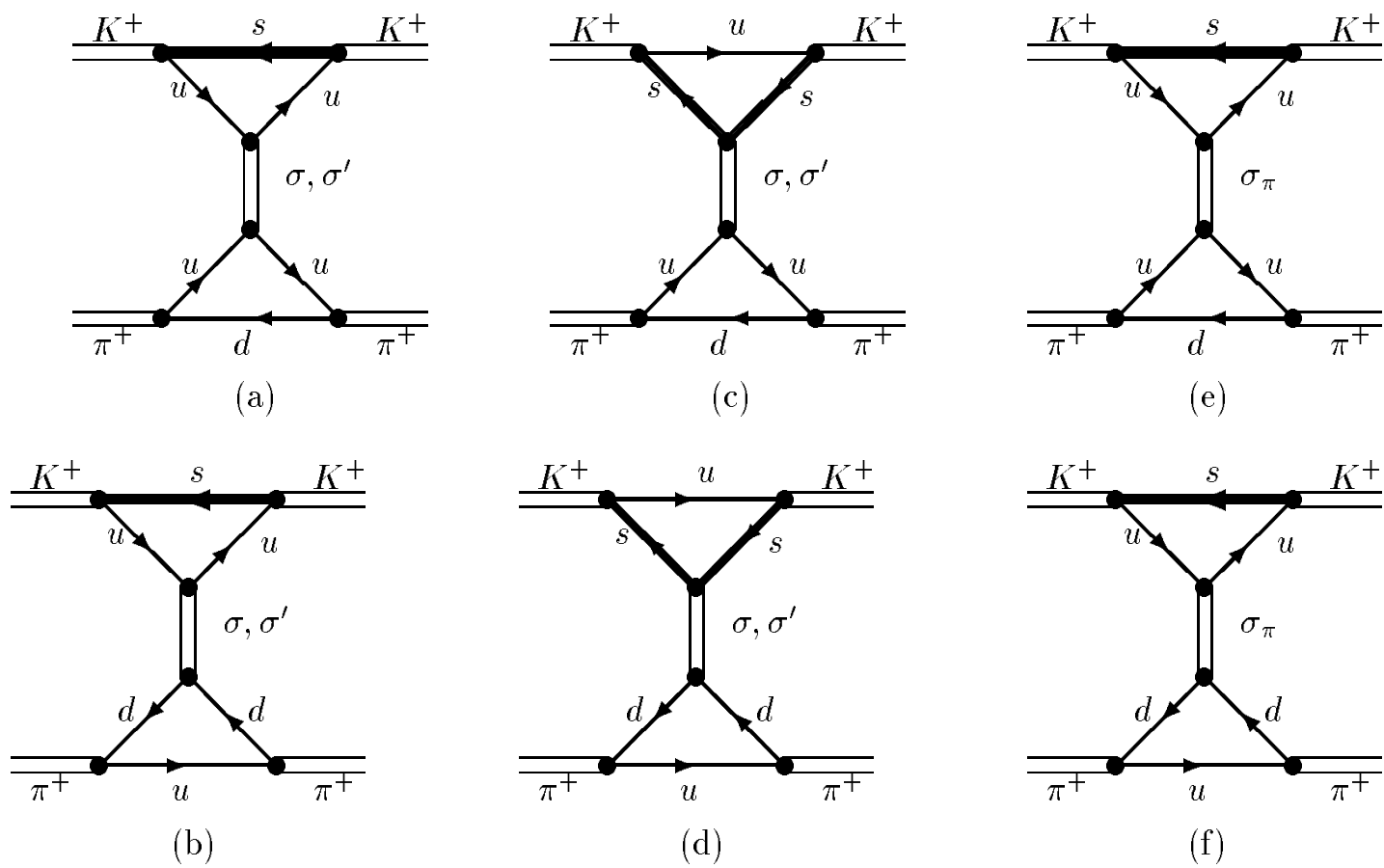
\begin{figure}
\caption[]{Possible $t$ channel meson exchange diagrams that
           contribute to the scattering $\pi^+ K^+\rightarrow \pi^+ K^+$.
Mesonic states are denoted by double lines, quarks by single lines.   The
$s$ quark is denoted by a heavy single line.}
\label{fig4}
\end{figure}

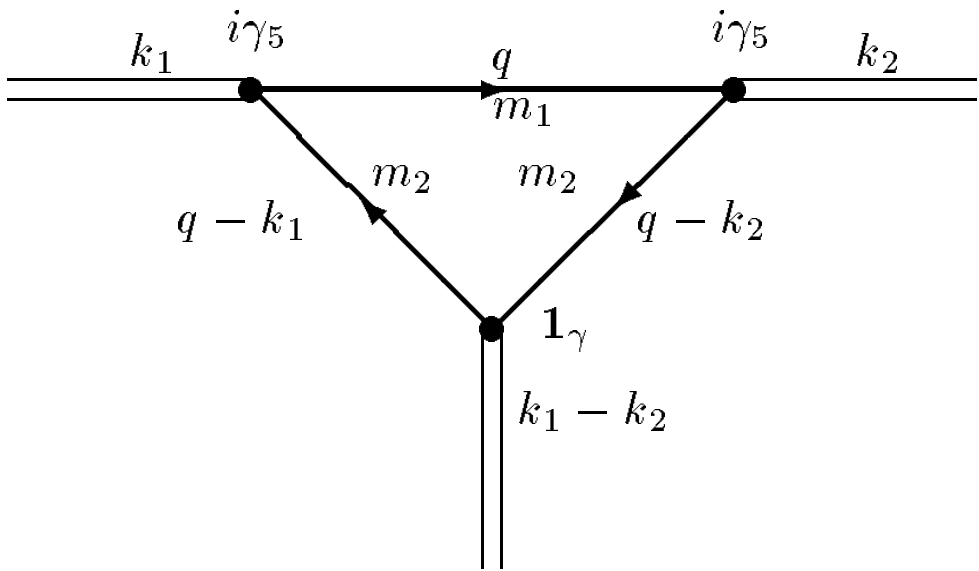
\begin{figure}
\caption[]{Generic three meson vertex function
           $\Gamma_1^{12}$ that occurs in the $t$ channel.
The external double lines denote meson states.   The internal single 
lines denote quarks with masses $m_1$ or $m_2$ as indicated.}
\label{fig5}
\end{figure}

As in the previous subsection, it is
necessary to analyse first which diagrams can contribute to the
$\pi^+K^+$ elastic scattering amplitude.   One notes immediately that
the $s$ channel graphs of Fig.~\ref{fig2} do not contribute, due to
charge conservation - the intermediate state would require a doubly
charged (scalar) meson, which does not exist within the model.  In the
$t$ channel, on the other hand, there are six possibilities.  These are
displayed explicitly 
in Fig.~\ref{fig4}.    Note that in all these diagrams, a
scalar meson is exchanged in the intermediate channel.    This is a
consequence of parity conservation and manifests itself directly in the
evaluation of the triangle diagrams that form a part of each amplitude
in Fig.~\ref{fig4}, in that the trace of the product of
$\gamma_5S\gamma_5S\gamma_5S$ vanishes, whereas that of $\gamma_5S\gamma_5
S1S$ does not.   We construct these diagrams by first considering the
triangle graph that is given in Fig.~\ref{fig5}.    From this Feynman
diagram, the three meson vertex is given as
\begin{eqnarray}
-i\Gamma^{12}_1(k_1,k_2) &=& - N_c\int\frac{d^4q}{(2\pi)^4}
{\rm tr}[i\gamma_5 iS^1(q)
i\gamma_5iS^2(q-k_1) iS^2(q-k_2)]
\label{e:mvertex} \\
&=& -N_c\int\frac{d^4q}{(2\pi)^4}
{\rm tr} \frac{i\gamma_5 i(\!\not\! q + m_1) i\gamma_5
i (\!\not\! q - \!\not\!
 k_1 + m_2) i(\!\not\! q-\!\not\! k_2 + m_2)}
{(q^2 - m_1^2)[(q-k_1)^2 - m_2^2][(q-k_2)^2 - m_2^2]} \nonumber
\quad ,
\end{eqnarray}
where the superscripts 1 and 2 denote the flavors of the internal
quarks and no trace over flavor has been included at this stage.
This must be combined with the effective propagators of the scalar mesonic 
sector.   We may simply use the form
\begin{equation}
D_{\sigma,\sigma'}(k_1-k_2) = M_{00}\lambda_0\times\lambda_0 +
M_{08}\lambda_0\times\lambda_8 +M_{80}\lambda_8\times\lambda_0 +
M_{88}\lambda_8\times\lambda_8  \quad ,
\label{e:dsigss}
\end{equation}
where  the $M_{ij}$ are a function of momenta, $M_{ij} =
M_{ij}(k_1-k_2)$.  If, for example, a strange quark couples to the
first vertex and a $u$ quark to the second of the mesonic amplitude, as
is the case in (c) and (d) of Fig.~\ref{fig4},  one can obtain the
flavor factors by considering the flavor components of
Eq.~(\ref{e:dsigss}) separately.  The 8-0 term of this expression, for
example, is thus
\begin{eqnarray}
\lambda_8(s,s) \cdot \lambda_0(u,u) &=& (0 0 1) \lambda_8
\left(\begin{array}{c} 0 \\0 \\ 1\end{array}\right) \cdot (1 0 0) \lambda_0
\left(\begin{array}{c} 1 \\0 \\ 0\end{array}\right) \nonumber \\
&=& -\frac{2\sqrt 2}{3} \quad ,
\label{e:flavor}
\end{eqnarray} 
and the other terms follow similarly, leading to the expression
\begin{equation}
D_{\sigma,\sigma'}^{qs} = \frac 23 M_{00} - \frac{\sqrt 2} 3 M_{08}
- \frac 23 M_{88} \quad .
\label{e:d1}
\end{equation}
This form of the intermediate mesonic scattering amplitude is required
for the evaluation of diagrams (c) and (d).    With it, one easily 
constructs the $t$ channel $\pi K$ scattering amplitudes in (c) and (d)
to be
\begin{equation}
iT^{3/2}_{t,\{(c) + (d)\}}(k_1,k_2,p_1,p_2) = -i2\cdot 4 g_{\pi q q }^2 g_{
Ksq}^2 \Gamma_1^{qs}(k_1,k_2) D_{\sigma,\sigma'}^{sq}(k_1-k_2) \Gamma_1^{qq}
(-p_2,-p_1) \quad ,
\label{e:tcd}
\end{equation}
since terms (c) and (d) are equal when $m_u=m_d$. In this expression, the
factor 4 arises from the external vertices, that each contribute a factor
$\sqrt2$. The factor 2 comes about since the contributions from graphs
(c) and (d) are equal.

In an analogous fashion, diagrams (a) and (b) can also be evaluated.
The intermediate mesonic scattering amplitude in this case is found to be
\begin{equation}
D_{\sigma,\sigma'}^{qq}(k_1-k_2) = \frac 23 M_{00} + \frac{2\sqrt 2}3 M_{08}
+\frac 13M_{88} \quad ,
\label{e:dqq}
\end{equation}
and in combination with Eq.~(\ref{e:mvertex}) 
leads to the expression
\begin{equation}
iT^{3/2}_{t,\{(a) + (b)\}} (k_1,k_2,p_1,p_2) = -i2\cdot 4 g_{\pi q q}^2
g_{Kqs}^2 \Gamma_1^{sq}(-k_2,-k_1)D_{\sigma,\sigma'}^{qq}(k_1-k_2)\Gamma_1
^{qq}(p_1,p_2)
\label{e:tab}
\end{equation}
for the $t$ channel $\pi^+ K^+$ scattering amplitudes depicted in (a) and (b).

There is no contribution from diagrams (e) and (f):   the only difference
between these two graphs is seen to be due to the flavor couplings to
the intermediate exchanged meson $\sigma_\pi$.   Since $\lambda_3(u,u)
= -\lambda_3 (d,d)$, the sum of these two terms cancel, and hence
\begin{equation}
iT^{3/2}_{t\{(e) + (f)\}} = 0 \quad .
\label{e:tef}
\end{equation}
Thus, in the $t$ channel, the $\pi^+K^+$ amplitude leads to 
\begin{equation}
  T^{3/2}_t = T^{3/2}_{t\{(a)+(b)\}} + T^{3/2}_{t\{(c)+(d)\} } \quad ,
\label{e:tchannel}
\end{equation}
with $T^{3/2}_{t\{(a)+(b)\}}$ and $T^{3/2}_{t\{(c)+(d)\} }$ as given
in Eqs.~(\ref{e:tab}) and (\ref{e:tcd}).

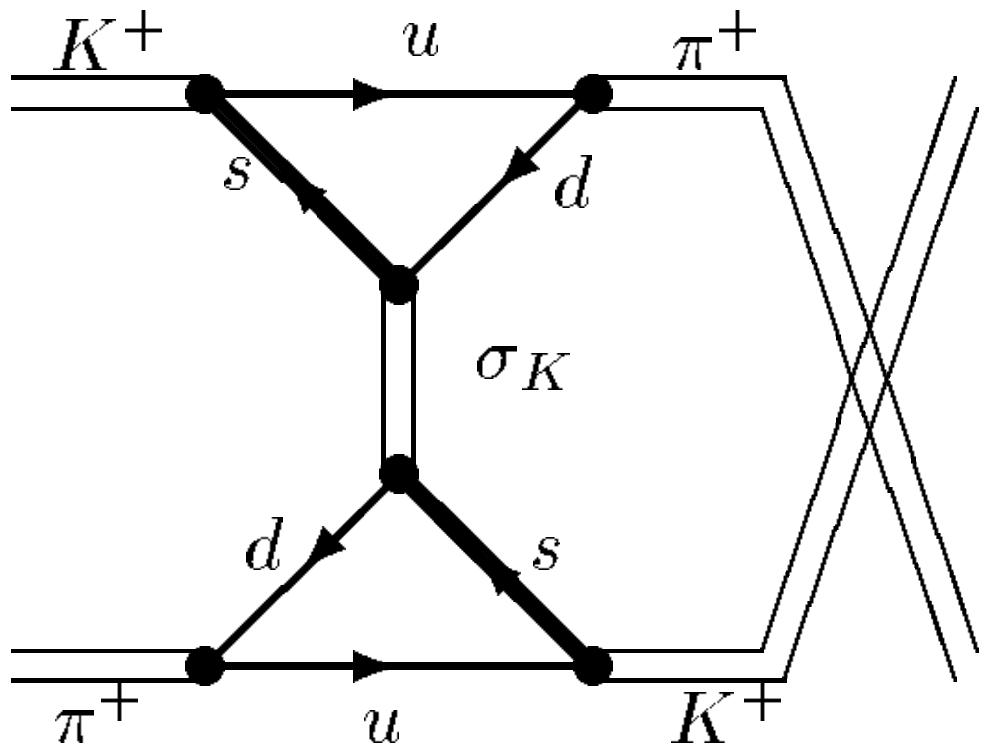
\begin{figure}
\caption[]{$u$ channel graph for the scattering amplitude
           $\pi^+ K^+ \rightarrow \pi^+ k^+$.
External double lines represent mesons, while single lines represent
quarks.  The $s$ quark is represented by a heavy single line.}
\label{fig6}
\end{figure}

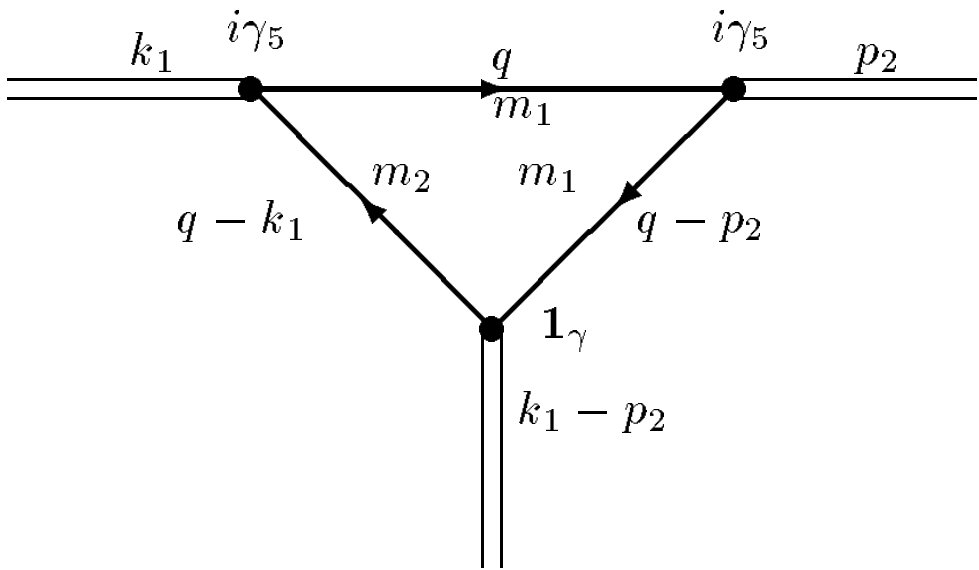
\begin{figure}
\caption[]{Generic three meson vertex function
           $\Gamma_2^{12}$ that occurs in the $u$ channel.
External double lines represent mesons, while single lines denote quarks
with masses $m_1$ or $m_2$ as indicated.}
\label{fig7}
\end{figure}

We now come to the $u$ channel graphs of Fig.~\ref{fig2}.   Such
processes require the exchange of an uncharged strange meson.   In the
NJL model, this corresponds to the exchange of a $\sigma_K$ and can be
realized within the model as depicted in Fig.~\ref{fig6}.
   An elementary vertex
needed for this process is shown in Fig.~\ref{fig7}.  Analytically, it
can be constructed as
\begin{eqnarray}
-i\Gamma_2^{12}(k_1,p_2) &=& -N_c\int\frac{d^4q}{(2\pi)^4}
{\rm tr}[i\gamma_5 iS^1(q)
i\gamma_5 iS^2(q-k_1) iS^1(q-p_2)] \label{e:mvertex2} \\
&=&-N_c\int\frac{d^4q}{(2\pi)^4}
{\rm tr}\frac{i\gamma_5 i(\!\not\! q + m_1) i\gamma_5 i(\!\not\!
q-\!\not\! k_1+m_2) i(\!\not\! q-\!\not\! p_2 + m_1)}
{(q^2-m_1^2)[(q-k_1)^2 - m_2)^2][(q-p_2)^2 + m_1^2]}
\nonumber
\end{eqnarray}
while the effective interaction mediated by the $\sigma_K$ meson follows
from Eq.~(\ref{e:mk}) as
\begin{equation}
D_{\sigma_K}(k_1-p_2) = \frac{2K_{44}^-}{1-2K_{44}^-\Pi_{sq}^{S}(k_1-p_2)}
\quad ,  \label{e:dsk}
\end{equation}
so that one arrives at the form
\begin{equation}
iT^{3/2}_u(k_1,k_2,p_1,p_2) = -8i g_{\pi q q}^2 g_{Ksq}^2
\Gamma_2^{qs}(k_1,p_2) D_{\sigma_K}(k_1-p_2)\Gamma_2^{qs}(k_2,p_1)
\label{e:tu}
\end{equation}
for the $u$ channel amplitude. 
The factor $8(={\sqrt 2}\ {}^6)$ in this expression comes from all flavor
factors
that contribute at each vertex.

The complete $I=3/2$ $\pi^+K^+\rightarrow \pi^+K^+$ scattering amplitude
is thus made up of three components,
\begin{equation}
T^{3/2}(k_1,k_2,p_1,p_2) = T^{3/2}_{{\rm box}}(k_1,k_2,p_1,p_2)
 + T^{3/2}_t(k_1,k_2,p_1,p_2) + T^{3/2}_u(k_1,k_2,p_1,p_2) \quad , 
\label{e:tfinal}
\end{equation}
with $T^{{\rm box}}$, $T^{3/2}_t$ and $T^{3/2}_u$ given by Eqs.~(\ref{e:box}),
(\ref{e:tchannel}) and (\ref{e:tu}) respectively.

\subsection{$\pi K$ scattering lengths}
The expressions given in the previous section enable one in principle
to calculate the $\pi K$ cross section for an arbitrary choice of
kinematic variables.  In practice, this turns out to be extremely
difficult, as the box diagram cannot easily be evaluated exactly for
arbitrary kinematics, requiring in general addition approximations.
In this paper, we restrict ourselves to a calculation of the scattering
lengths, which simplifies the calculation somewhat.     The kinematic
threshold is given by
\begin{equation}
s = (m_\pi + m_K)^2
\label{e:skin}
\end{equation}
with 
\begin{equation}
t=0,\quad u=(m_\pi - m_K)^2, \quad q=0 \quad .
\label{e:tkin}
\end{equation}
These conditions can be fulfilled by choosing
\begin{equation}
k_1=k_2 = k = \left(\begin{array}{c} m_K \\ \vec 0 \end{array}\right)
\label{e:kkin}
\end{equation}
and
\begin{equation}
p_1=p_2=p=\left(\begin{array}{c}  m_\pi \\ \vec 0 \end{array}\right) 
\quad .
\label{e:pkin}
\end{equation}
Using these kinematics, the box term, as well as the triangle and 
intermediate meson exchange graphs making up the $t$ and $u$ channel
contributions, can be expressed in terms of the ``elementary'' integrals
\begin{eqnarray}
F^i &=& \int\frac{d^4q}{(2\pi)^4} \frac 1{q^2-m_i^2} 
\label{e:ffunc}\\
M^{12}(p) &=& \int\frac{d^4q}{(2\pi)^4} \frac 1{(q^2-m_1^2)[(q-p)^2 - m_2^2]}
\label{e:mfunc}\\
N^{12}(p) &=& \int\frac{d^4q}{(2\pi)^4} \frac 1{(q^2-m_1)^2[(q-p)^2-m_2^2]}
\label{e:nfunc}\\
P^{12}(p,k) &=& \int\frac{d^4q}{(2\pi)^4}\frac 1{(q^2-m_1^2)[(q-p)^2 -m_1^2]
[(q-k)^2 -m_2^2]}
\label{e:pfunc} \\
Q^{12}(p,k) &=& \int\frac{d^4q}{(2\pi)^4} \frac 1{(q^2-m_1^2)^2[(q-p)^2
-m_1^2][(q-k)^2-m_2^2]}
\label{e:qfunc}
\end{eqnarray}
that form the building blocks for these functions.   We list the results for
the components of the $T^{3/2}$ amplitude for the threshold kinematics.

\subsubsection{Box diagram}  The kinematic choice of Eqs.~(\ref{e:kkin}) and
(\ref{e:pkin}) lead to the forms
\begin{eqnarray}
&&iT^{3/2}_{{\rm box}}(k,k,p,p) = -4N_cg_{\pi q q}^2 g_{Kqs}^2
\label{e:tboxkin} \\
&&\times \int\frac{d^4q}{(2\pi)^4}
\left\{\frac {{\rm tr}\left[\gamma_5(\!\not\! q - \!\not\! k + 
m_s)\gamma_5(\!\not\! q + m_q)\gamma_5(\!\not\! q -\!\not\!
 p + m_q)\gamma_5(\!\not\! q + m
_q)\right]}{[(q-k)^2 - m_s^2][q^2-m_q^2]^2[(q-p)^2-m_q^2]} \right\}
\quad . \nonumber
\end{eqnarray}
After some calculation along the lines that will be indicated for the
three meson vertex function to follow, one finds
\begin{eqnarray}
iT^{3/2}_{{\rm box}}(k,k,p,p) =&-& 8 g_{\pi q q}^2 g_{Kqs}^2 N_c
\left\{M^{qq}(0) +M^{qs}(k-p)-p^2N^{qq}(p) \right. \nonumber\\
&-& \left[k^2-(m_q-m_s)^2\right]N^{qs}(k) - 2k\cdot p P^{qs}(p,k) 
\nonumber \\
&+& \left. p^2\left[k^2 - (m_q-m_s)^2\right] Q^{qs}(p,k) \right \} \quad
.
\label{e:tboxele}
\end{eqnarray}

\subsubsection{Meson exchange graphs}    The three meson vertex functions
are required for the $t$ and $u$ channel graphs.    From
Eq.~(\ref{e:mvertex}), one has
\begin{equation}
-i\Gamma_1^{12}(k,k) = -iN_c\int\frac{d^4q}{(2\pi)^4} \frac{ {\rm tr}
\left[\gamma_5( \!\not\! q + \!\not\! k + m_1)\gamma_5(\!\not\!
 q + m_2)(\!\not\! q + m_2)
\right]} {[q^2-m_2^2]^2[(q+k)^2-m_1^2]} \quad ,
\label{e:gam1}
\end{equation}
which reduces to 
\begin{equation}
-i\Gamma_1^{12}(k,k) = -4iN_c\int \frac{d^4q}{(2\pi)^4}
\frac{q^2(m_1-2m_2) - 2m_2q\cdot k + m_1m_2^2}
{(q^2-m_2^2)^2[(q+k)^2 -m_1^2]} \quad , 
\label{e:gam1ele}
\end{equation}
after performing the trace.   Using the relation
\begin{equation}
2k\cdot q = [(k+q)^2 - m_1^2] - [q^2-m_2^2] - k^2 - m_2^2 + m_1^2
\label{e:rel}
\end{equation}
leads one to the expression
\begin{eqnarray}
-i\Gamma_1^{12}(k,k) &=& -4iN_c\Big\{-m_2M^{22}(0) + (m_1-m_2)M^{21}(k)
\nonumber \\
& & \hspace{1cm} + m_2[k^2- (m_2-m_1)^2]N^{21}(k)\Big\}
\label{e:gam1elem}
\end{eqnarray}
in terms of the elementary integrals. In a similar fashion, the second
vertex function that is required for the $u$ channel $T$ matrix
amplitude of Eq.~(\ref{e:mvertex2}) and which was shown in
Fig.~\ref{fig7} can be decomposed as
\begin{eqnarray}
-i\Gamma_2^{12}(k,p) 
&=& 2N_c i\Big\{(m_2-m_1) M^{12}(k) + (m_2+m_1)M^{12}(k-p) \nonumber \\
&&\quad - [m_2p^2 + m_1k^2-m_1(p-k)^2]P^{12}(p,k)\Big\} \quad .
\label{e:mvertex2ele}
\end{eqnarray}
In addition to this, the scalar polarization function
is given as
\begin{equation}
-i\Pi_S^{12}(k) = -2N_c\int\frac{d^4q}{(2\pi)^4}
\frac{{\rm tr}[i(\!\not\! q + m_1)i(
\!\not\! q - \!\not\! k + m_2)]}{[q^2-m_1^2][(q-k)^2-m_2^2]}
\quad , \label{e:pols}
\end{equation}
and is expressed as
\begin{equation}
-i\Pi_S^{12}(k) = 4N_c\{F^1+F^2+[(m_1+m_2)^2 - k^2]M^{12}(k) \}
\quad . \label{e:polsele}
\end{equation}
Note that the pseudoscalar polarization differs from this function only in
the
replacement $(m_1+m_2)^2\rightarrow (m_1-m_2)^2$.

The complete expression for the $I=3/2$ scattering amplitude at the 
kinematic threshold follows now on constructing
\begin{equation}
iT^{3/2}_{t\{(a)+(b)\}}(k,k,p,p) = -8i g_{\pi q q}^2 g_{Kqs}^2
\Gamma_1^{sq}(k,k)\left(\frac23 M_{00} + \frac{2\sqrt 2}3 M_{08} + \frac 13
M_{88}\right)\Gamma_1^{qq}(p,p) 
\label{e:tabthreshold}
\end{equation}
and
\begin{equation}
iT^{3/2}_{t\{(c)+(d)\}}(k,k,p,p) = -8ig_{\pi qq}^2 g_{Kqs}^2
\Gamma_1^{qs}(k,k)\left(\frac 23M_{00} - \frac{\sqrt 2}3 M_{08} - \frac 23
M_{88}\right)\Gamma_1^{qq}(p,p)
\label{e:tcdthreshold}
\end{equation}
for the $t$ channel, and
\begin{equation}
iT^{3/2}_u(k,k,p,p) = -8ig_{\pi q q}^2 g_{Kqs}^2 \left[ \Gamma_2^{qs}(k,p)
\right]^2 \frac{2K_{44}^-}{1-2K_{44}^-\Pi_S^{qs}(k-p)}
\label{e:tuthreshold}
\end{equation}
for the $u$ channel, with the couplings
\begin{eqnarray}
g_{Kqs}^{-2} &=& - 4iN_c
\Bigg\{M^{qs}(k) + \frac{(m_q-m_s)^2-k^2}{2k^2}\Big[M^{qs}(k) - M^{ss}(0)
\nonumber \\
& & \hspace{2cm} + (m_s^2-m_q^2 +k^2)N^{sq}(k)\Big]\Bigg\}_{k^2=m_K^2}
\label{e:gkthreshold}
\end{eqnarray}
and
\begin{equation}
g_{\pi qq}^{-2} = -2iN_c\left[M^{qq}(0) + M^{qq}(k)
- k^2N^{qq}(k)\right]_{k^2=m_\pi^2},
\label{e:gpithreshold}
\end{equation}
which can be derived from the definitions in 
Eq.~(\ref{e:gs}).   Then one has
\begin{eqnarray}
T^{3/2}(k,k,p,p) &=& T_{{\rm box}}^{3/2}(k,k,p,p)+ T^{3/2}_{t\{(a)+(b)\}}
(k,k,p,p) \nonumber \\
&+& T^{3/2}_{t\{(c)+(d)\}}(k,k,p,p) + T^{3/2}_u(k,k,p,p)
\quad . \label{e:tthreshold}
\end{eqnarray}
The scattering lengths are evaluated as
\begin{equation}
a^I = \frac1{8\pi(m_\pi + m_K)} \Re\left[ T^I(s=(m_\pi+m_K)^2,t=0,
u=(m_\pi-m_K)^2)\right]
\quad , \label{e:scattl}
\end{equation}
so that $a^{3/2}$ follows directly from Eq.~(\ref{e:tthreshold}), while
$a^{1/2}$ is obtained  subsequently through the relation
Eq.~(\ref{e:relation}).

\section{Numerical evaluation of the $\pi K$ scattering lengths}

\subsection{T=0 Case}
\label{sec4a}
In order to evaluate the scattering lengths, the functions $F$, $M$,
$N$, $P$ and $Q$ of Eqs.~(\ref{e:ffunc}) - (\ref{e:qfunc}) must be
evaluated.  Since the NJL model is non-renormalizable, a cutoff
parameter must be introduced to regulate these functions.   This can be
done in a covariant fashion using the Pauli-Villars scheme, or
non-covariantly by introducing a cutoff on the three momentum $|\vec
p|< \Lambda_3$.   For consistency, a cutoff is
included in all integrals, including $N$, $P$ and $Q$, that are convergent.
 Technical details for the Pauli-Villars integrals are
given in Appendix \ref{app:pv}. The $O(3)$ calculation is performed
directly numerically.    Our first evaluation is performed at $T=0$.

In Table~\ref{table1}, we list two sets of parameters for the
Pauli-Villars calculation that were taken from Ref.~\cite{bv},
 as well as two
sets of parameters for an O(3) cutoff, taken from Refs.~\cite{hatkun,pr}.
    These
calculations are all non-chiral,  all assuming  non-vanishing values
for the current quark masses.    Using these parameter sets, we obtain 
values for the pseudoscalar meson sector at $T=0$ that are listed in 
Table~\ref{table2}.   The experimental values are also given.   The scalar
meson masses for $T=0$ for the corresponding parameter sets are given
in Table~\ref{table3}.    Using these values, the $s$-wave scattering
lengths $a_0^{1/2}$ and $a_0^{3/2}$ are calculated at $T=0$.
   The results for
 $a_0^{3/2}$, obtained from the box,
$t$ and $u$ channels for each parameter set, are given in Table~\ref{table4},
 with those for
$a_0^{1/2}$ given in Table~\ref{table5}.   The final results
for the scattering lengths arise  from a delicate cancellation
of the box, $t$ and $u$ channel values, which is  evident from these
tables\footnote{To obtain a deeper understanding of this fact, it is
necessary to perform a chiral expansion of the box, $s$, $t$ and $u$ channel
graphs in the two variables ($m_\pi/\Lambda$
 and $m_K/\Lambda$) analytically along the lines suggested by
Refs\cite{bandco} and \cite{schulze}. This is a difficult problem in its own 
right and will not be addressed here.}.    In the model, we find 
 that the calculated values of $a_0^{3/2}$ range
from $-0.003$ to $-0.04$, while that of $a_0^{1/2}$ range from $0.13$ to
$0.18$, in units of $m_\pi$.    Our results for $a_0^{1/2}$ fall well within
the experimentally known range of $0.13$ to $0.24$, while those for
$a_0^{3/2}$ are lower, the experimental values being $-0.13$ to $-0.05$.   
Standard chiral perturbation theory
gives $a_0^{1/2} = 0.17$ and $a_0^{3/2}=-0.05$ \cite{bkm1} for these values.
Thus our $T=0$ values, while dependent on the regularization scheme, can be
considered reasonable.

\subsection{$\pi K$ scattering lengths for $T\ne 0$} \label{sec4}
At finite values of the temperature, the $\pi K$ scattering graphs can be 
analysed as was done at $T=0$, but within the imaginary time or Matsubara
formalism.   This corresponds to making the formal substitution
\begin{equation}
\int\frac{d^4q}{(2\pi)^4} \rightarrow \frac i\beta\sum_n\int_\Lambda\frac
{d^3q}{(2\pi)^3}
\label{e:subst}
\end{equation}
and replacing the fermionic (bosonic) $q_0$ by the discrete frequencies
$i\omega_n$, $\omega_n = (2n+1)\pi/\beta$ or $\omega_n=2n\pi/\beta$, $n=
0,\pm1,\pm2,\pm3\cdots$ respectively.  The final expressions are given 
in Appendix B.   As is indicated in Eq.~(\ref{e:subst}), an O(3) cutoff is
naturally implemented.

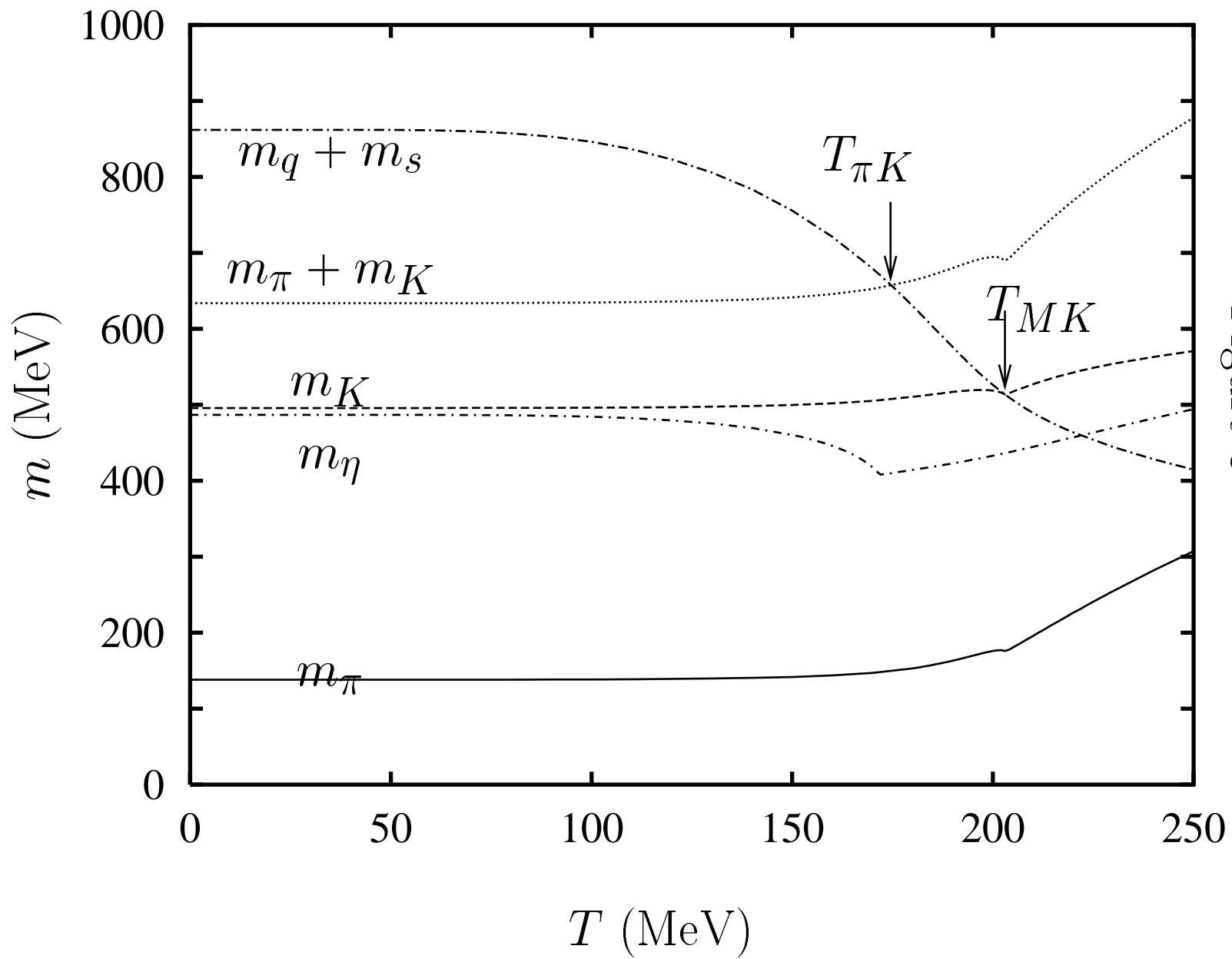
\begin{figure}
\caption[]{Pseudoscalar meson masses versus temperature
 for the parameter set labeled HK of Table~\ref{table1}.  Shown are
$m_\pi$  (solid line), $m_K$ 
           (dashed line), and $m_\eta$ (lower dot-dashed line).
The combinations  $m_\pi+m_K$ (dotted  line) and  $m_q+m_s$ (upper
dot-dashed line) are also shown.   $T_{\pi K}$ and $T_{MK}$ are indicated
with arrows.}
\label{fig8}
\end{figure}

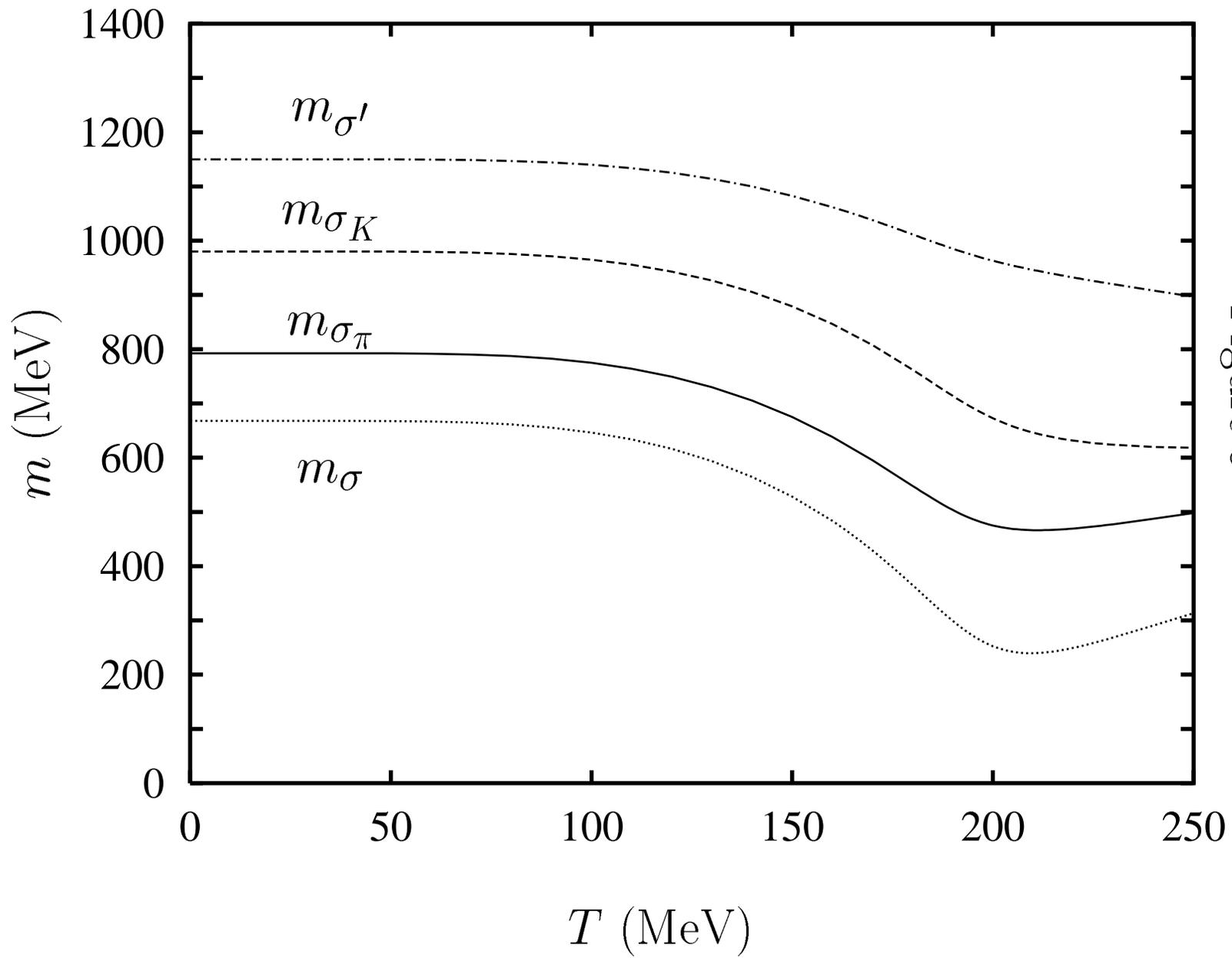
\begin{figure}
\caption[]{Scalar meson masses versus temperature for the parameter
set labeled HK.   Shown are $m_\sigma$ (dotted line), $m_{\sigma_\pi}$
(solid line), $m_{\sigma_K}$ (dashed line) and $m_{\sigma'}$ (dot-dashed
line).}
\label{fig9}
\end{figure}

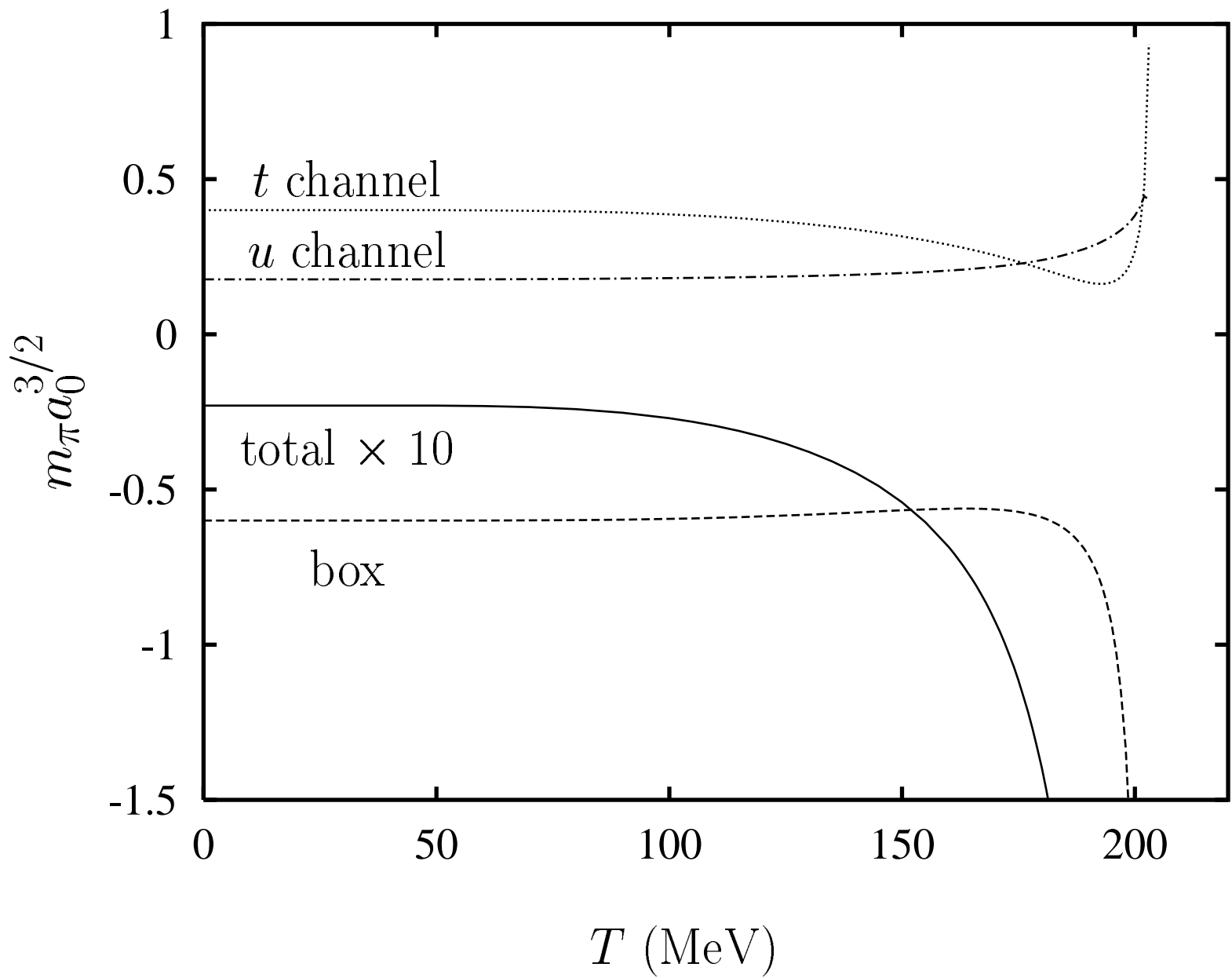
\begin{figure}
\caption[]{Temperature dependence of $a_0^{3/2}$ as well as the
decomposition of  the $t$, $u$ and box graph
	   contributions to this quantity for the parameter set of HK,
 in units of $m_\pi$.   The full curve for $a_0^{3/2}$ is indicated by
the solid line, the box diagram contribution by dashed lines, the $t$ and
$u$ contributions by a dotted and dot-dashed curve respectively.}
\label{fig10}
\end{figure}

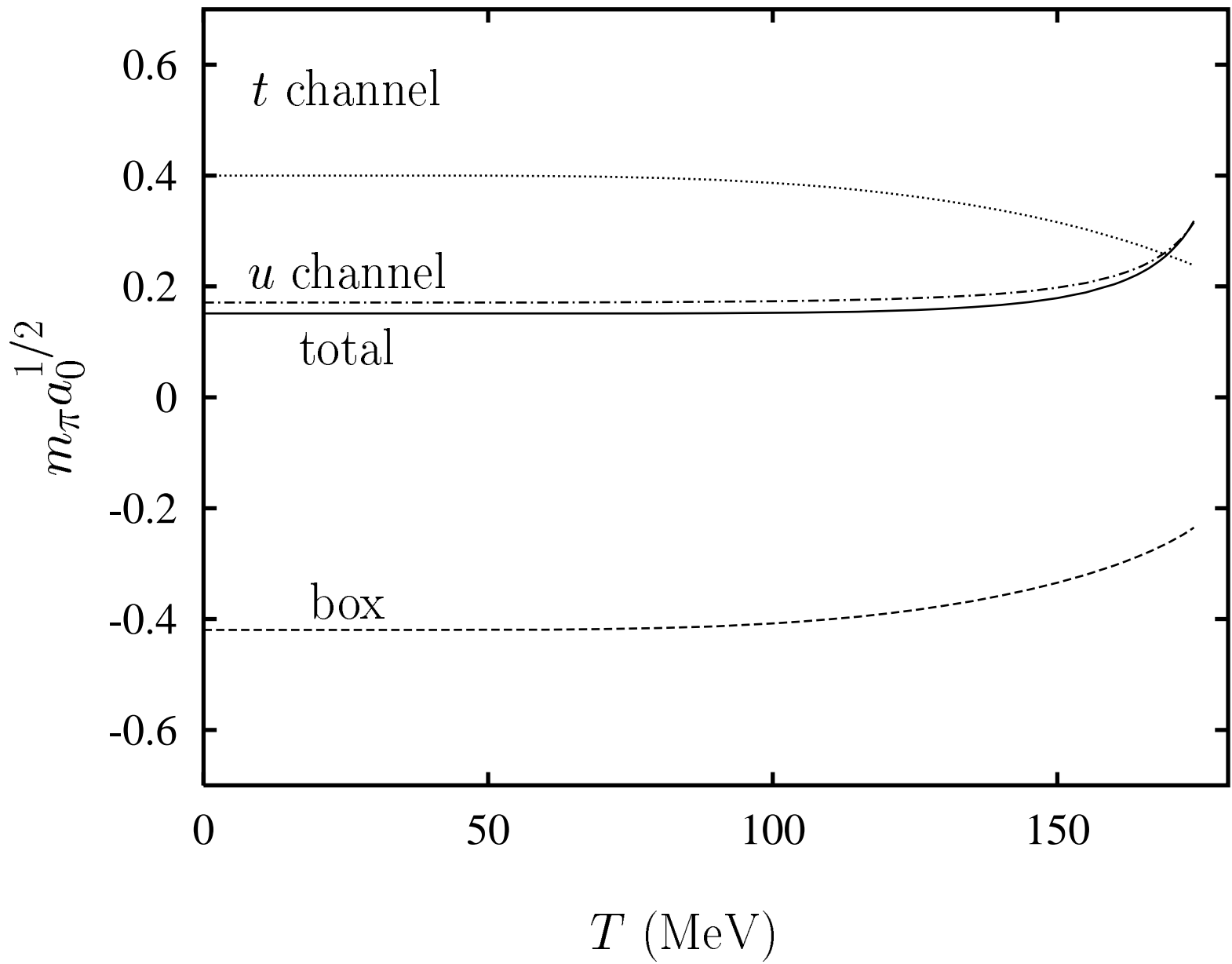
\begin{figure}
\caption[]{Temperature dependence of $a_0^{1/2}$ as well as 
the $t$, $u$ and box graph
	   contributions to this quantity for the parameter set of HK and 
in units of $m_\pi$. The full curve for $a_0^{1/2}$ is indicated by the 
solid line, the box diagram contribution by dashed lines, the $t$ and
$u$ contributions by a dotted and dot-dashed curve respectively.}
\label{fig11}
\end{figure}

The pseudoscalar meson pole masses vary with temperature, and this variation
is shown in  Fig.~\ref{fig8},
 together with the dissociation threshold for the
kaon, $m_q+ m_s$, which leads to a kaon dissociation temperature
$T_{MK}$. The function $m_K + m_\pi$ is also shown. The temperature
at which this function crosses $m_q+ m_s$ defines the point $T_{\pi
K}$ at which a meson pair can dissociate into a quark pair.
Numerically, for the set of parameters denoted as HK in
 Table~\ref{table1},
 we have $T_{\pi K}=174.5$~MeV, while $T_{MK}= 203.0$~MeV.
We also show the temperature dependence of the 
 scalar mesonic sector  in Fig.~\ref{fig9} for this parameter set.
Note that the behavior of the $\sigma_\pi$ meson with temperature
\cite{pr} has been confirmed qualitatively in lattice gauge calculations
\cite{laer2}.
Using the temperature dependent  functions for the mesons,
the temperature dependence of the box, $t$ and $u$ channel amplitudes
is evaluated and used to construct the scattering lengths.   In Fig.~\ref
{fig10}, we show the individual contributions from these channels, together
with the evaluated scattering length $a_0^{3/2}$.   One sees that
all graphs
diverge close to the Mott point. The box and $u$ channel graphs do not
change substantially up until the Mott point, while the $t$ channel meson
exchange diagram displays a somewhat faster variation with temperature.
Consequently, the resulting change in $a_0^{3/2}$ is noticeable - the
change, within this model calculation, is quantitatively given as
 $a_0^{3/2}(T=150$~MeV$)/a_0^{3/2}(T=0) = 2.36$, in contrast to the
completely flat temperature dependence found for $\pi\pi$ scattering
lengths \cite{qufamily}.
The temperature dependence of the $a_0^{1/2}$ is less dramatic:
Fig.~\ref{fig11} shows that the sum of the functional components lead to
a flat behavior.    Quantitatively, one has
$a_0^{1/2}(T=150$~MeV$)/a_0^{1/2}(T=0)=1.18$.

\section{Summary and Conclusions} \label{sec5}
In this paper, we have evaluated the temperature dependence of the $\pi
K$ scattering lengths $a_0^{3/2}$ and $a_0^{1/2}$ within the framework
of the SU(3) NJL model. Ordering terms according to an expansion in
$1/N_c$ is used to determine the leading set of diagrams that
contribute to the scattering amplitude.
One finds a box-like graph, as well as $t$ and $u$ channel graphs.   To
this order in $N_c$, no unitarity corrections are present.    Because
of this, only the scattering lengths are investigated, although the
formulae are given for arbitrary kinematics, so that
the scattering amplitude can be evaluated for arbitrary kinematics.    The
restriction here to threshold kinematics also simplifies  the computation
considerably.

At $T=0$, we have 
calculated 
 the $l=0$ scattering
lengths  $a_0^{1/2}$ and  $a_0^{3/2}$ 
for different regularization schemes, that of Pauli and Villars, and also
using an O(3) cutoff.    We have simply used parameters sets of other
authors and have not fitted any additional quantities.   We find
that $a_0^{1/2}$  falls well within the experimentally known data bracket,
while $a_0^{3/2}$  is slightly lower than the bracketing 
 values given experimentally.
Unfortunately the experimental values are not known precisely at this
stage.

In investigating the finite temperature behavior of the scattering
amplitudes,
we observe that the 
 box and $u$ channel graphs, the latter of which occurs by resonance
exchange of a $\sigma_K$ scalar meson, do not vary strongly with the
temperature.  However, variation of the $t$ channel amplitude that is
controlled by the exchange of $\sigma$ and $\sigma'$ mesons displays a
slightly more pronounced temperature dependence, leading to a
temperature change of $a_0^{3/2}$ of a factor of about 2  over 150~MeV. 
This is in contrast to the case for $\pi\pi$ scattering, in which
the $\pi\pi$ scattering lengths are found to be essentially independent of
temperature over a range of about 150 MeV.    The study of kaon physics
appears here to give a stronger functional dependence than in the pionic
case.   Since this is driven essentially by the temperature dependence of
the intermediate scalar mesons, it seems that processes
that involve the exchange of
mesons in the scalar sector may ultimately provide us with a clean
signal for a chiral phase transition.

\section*{ACKNOWLEDGMENTS}
This work has been supported in part by the Deutsche
Forschungsgemeinschaft DFG under the contract number Hu 233/4-4, and by
the German Ministry for Education and Research (BMBF) under contract
number 06 HD 742.  One of us (S.P.K.) would like to thank S. Kahana for
the generous hospitality at Brookhaven National Laboratory, where this
paper was written.

\appendix

\section{Pauli-Villars Regularization}\label{app:pv}
Standard Feynman parametrization \cite{nak},
\begin{equation}
\prod_{i=1}^n\frac 1{A_i^{\zeta_i}} = \frac{\Gamma(\zeta)}{\prod_{i=1}^n
\Gamma(\zeta_i)}\int_0^1\left( \prod_{i=1}^n dx_i\, x_i^{\zeta_i -1}
\right) 
\frac{\delta(1-x)}{(\sum_{i=1}^n x_iA_i)^\zeta}
\quad ,  \label{a:1}
\end{equation}
with $\zeta = \sum_{i=1}^n\zeta_i$ and $x=\sum_{i=1}^n x_i$ is used
for the integrals Eqs.~(\ref{e:ffunc})--(\ref{e:qfunc}).   All calculations
are performed at the kinematic threshold.    One has
\begin{equation}
F = \int\frac{d^4 q}{(2\pi)^4} \frac 1{q^2 - m^2}
\quad ,  \label{a:2}
\end{equation}
which, after Wick rotation, gives
\begin{equation}
F= - \frac{i}{(4\pi)^2} \left[ q^2 - m^2\ln|q^2+m^2|\right]_0^\infty
\quad . \label{a:3}
\end{equation}
The Pauli-Villars regularization scheme is then implemented:
\begin{eqnarray}
F &=& - \frac{i}{(4\pi)^2}\sum_{j=0}^2 C_j
\left[(m^2+\alpha_j\Lambda^2)\ln
|q^2+m^2+\alpha_j\Lambda^2|\right]_0^\infty \nonumber \\
 &=& - \frac{im^2}{(4\pi)^2} \left[
\ln\left|1 - \left(\frac{m^2}{\Lambda^2}+1\right)^{-2}\right|
+ 2\frac{\Lambda^2}{m^2}
\ln\left|1+\left(\frac{m^2}{\Lambda^2}+1\right)^{-1}\right|\right]
\quad , \label{a:4}
\end{eqnarray}
where $m_j=m^2+\alpha_j\Lambda^2$ and the standard set of parameters
$C_0 =1$, $C_1=1$, $C_2=-2$, $\alpha_0=0$, $\alpha_1= 2$ and $\alpha_2=1$
are used.

The function $M^{12}(p)$ from Eq.~(\ref{e:mfunc}) can be integrated after Wick
rotation.   One has
\begin{eqnarray}
M^{12}(p) &=& i\int\frac{d^4q_E}{(2\pi)^4} \int_0^1 dx\frac 1{(q^2 + R^2)^2} 
\nonumber \\
&=& \frac{i}{(4\pi)^2}\int_0^1 dx  \left[ \frac {R^2}{q^2+R^2} +
\ln(q^2+R^2)\right]_0^\infty 
\label{a:5}
\end{eqnarray}
with $R^2 = m_1^2x + m_2^2(1-x) - p^2(1-x)x$.   The Pauli-Villars 
regularized version reads
\begin{equation}
M^{12}(p) = \frac{i}{(4\pi)^2}\sum_{j=0}^2 C_j \int_0^1 dx
\left [\frac {R_j^2}{q^2+R_j^2} + \ln (q^2+R_j^2)\right]_0^
\infty
\quad , \label{a:6}
\end{equation}
with the replacements $m_1^2\rightarrow m_{1j}^2 = m_1^2 + \alpha_j
\Lambda^2$ and $m_2^2\rightarrow m_{2j}^2=m_2^2 + \alpha_j\Lambda^2$ in
$R^2$ defining $R_j^2$. One obtains
\begin{equation}
M^{12}(p) = -\frac{i}{(4\pi)^2} \sum_{j=0}^2 C_j \int_0^1 dx \ln
\left[\frac {(m_1^2 - m_2^2)x + m_{2j}^2 -p^2(1-x)x}{m_2^2} \right ]
\quad . \label{a:7}
\end{equation}
The $x$-integration can be performed explicitly leading to
\begin{equation}
M^{12}(p) = -\frac{i}{(4\pi)^2}\sum_{j=0}^2C_j\left[\ln\frac {m_{2j}^2}{m_2^2}
+ \frac{p^2+m_1^2-m_2^2}{2p^2} \ln\left(\frac{m_{1j}^2}{m_{2j}^2}\right)
+ \frac{1}{p^2}A_j\right]
\quad ,  \label{e:a8}
\end{equation}
with $\Delta_j = (m_1^2-m_2^2-p^2)^2 - 4m_{2j}^2p^2$ and
\begin{equation}
A_j = \sqrt{-\Delta_j}\left[
\tan^{-1}\left(\frac{p^2+m_1^2-m_2^2}{\sqrt{-\Delta_j}}\right)
- \tan^{-1}\left(\frac{-p^2 + m_1^2 -m_2^2}{\sqrt{-\Delta_j}}\right)\right]
\label{e:a9}
\end{equation}
if $\Delta_j < 0$ and
\begin{equation}
A_j = \sqrt{\Delta_j}\left[\frac{1}{2}\ln
\frac{p^2-m_{1j}^2-m_{2j}^2 + \sqrt{\Delta_j}}
     {p^2-m_{1j}^2-m_{2j}^2 - \sqrt{\Delta_j}}
-i\pi\Theta\left(p^2-(m_1-m_2)^2\right)\right]
\end{equation}
if $\Delta_j>0$.
Following the same prescription for $N^{12}(p)$ leads to
\begin{eqnarray}
N^{12}(p) &=& -\frac i{(4\pi)^2}\int_0^1 dx\frac x{R^2} \nonumber \\
&=& - \frac i{(4\pi)^2}\frac 1{2p^2} \left[\ln\left(\frac{m_1^2}{m_2^2}\right) 
+2\frac {m_1^2 - m_2^2 -p^2}{\Delta} A\right]
\quad . \label{a:10}
\end{eqnarray}

Both $P^{12}(p,k)$ and $Q^{12}(p,k)$ of Eqs.~(\ref{e:pfunc}) and
(\ref{e:qfunc}) are functions of two variables and are somewhat more
complicated.  Feynman parametrization followed by a Wick rotation and
integration over the momentum variable leads to
\begin{equation}
P^{12}(p,k) = -\frac i{(4\pi)^2}\int_0^1 dx \int_0^1 dy \frac y{\hat R^2}
\label{a:11}
\end{equation}
where
\begin{equation}
\hat R^2 = m_1^2(1-y+xy) + m_2^2(1-x)y - [k^2(1-x) + p^2x]y(1-y) - x(1-x) y^2u
\quad , \label{a:12}
\end{equation}
with $u=(p-k)^2$.    The integral in $x$ can again be performed.   One
finds
\begin{equation}
P^{12}(p,k) = -\frac{2i}{(4\pi)^2}\int_0^1
\frac{dy}{\sqrt{\hat\Delta}}\left[\tan^{-1}
\left(\frac{2uy +\rho - \kappa y}{\sqrt{\hat\Delta}}\right)
- \tan^{-1}\left(\frac
{\rho - \kappa y}{\sqrt{\hat\Delta}}\right)\right]
\quad . \label{a:13}
\end{equation}
where the discriminant in this case is
\begin{eqnarray}
-y^2\hat\Delta &=& (\rho y - \kappa y^2)^2 - 4(m_1^2 + \omega y + k^2y^2)y^2u 
\nonumber \\
&=& -y^2\left[(4m_1^2u-\rho^2) + (2\rho \kappa + 4\omega u)y - (\kappa^2
-4k^2u)y^2\right]
\quad , \label{a:14}
\end{eqnarray}
with $\rho = m_1^2 - m_2^2 + k^2 - p^2$, $\kappa = k^2 + u - p^2$ and
$\omega = m_2^2 - m_1^2 - k^2$.

A similar analysis leads to
\begin{equation}
Q^{12}(p,k) = \frac{i}{(4\pi)^2}\int_0^1\int_0^1
dx dy\frac{y(1-y)}{\left(\hat R^2\right)^2}
\quad , \label{a:15}
\end{equation}
in which the $x$ integration can once more be carried out with the
aid of the integral
\begin{equation}
\int_0^1\frac {dx}{[ax^2+bx+c]^2} = -\frac{1}{\Delta}
\left[\frac{2ax+b}{ax^2+bx+c}
+ \frac{4a}{\sqrt{-\Delta}}\tan^{-1}\left(\frac{2ax+b}{\sqrt
{-\Delta}}\right)\right]_0^1
\label{a:16}
\end{equation}
with $\Delta=b^2-4ac$.
Note that the Pauli-Villars regularization procedure has not been
carried out explicitly on $N^{12}(p)$, $P^{12}(k,p)$ and $Q^{12}(k,p)$,
which are convergent.  For consistency within the model, however, this
should be done.    The implementation is simple.   The standard
replacement of each integral by a sum of three terms with $C_j$ and
$m_{ji}$ is constructed and evaluated numerically.   The forms for
$P$ and $Q$, Eqs.(\ref{a:13}) and (\ref{a:16}) are valid in the case that
the integrands are non-singular.

\section{Finite temperature integrals}
Using the abbreviation $E_j^2 = \vec q^2 + m_j^2$ for the energy of
particle with flavor $j$,
 one finds the following forms for the integrals
Eqs.~(\ref{e:ffunc}) -- (\ref{e:qfunc}) at the kinematic threshold:
\begin{equation}
F^j = -\frac{i}{4\pi^2}\int_0^\Lambda dq \frac {q^2}{E_j}
\tanh\left(\frac{\beta E_j}{2}\right)
\label{b:1}
\end{equation}
while
\begin{eqnarray}
M^{12}(p) &=& - \frac{i}{16\pi^2}\frac{1}{p_0^2}\Bigg\{
{\cal P} \!\!\! \int dq \frac{q^2}{q_0^2-q^2}\Bigg[
\tanh\left(\frac{\beta E_1}{2}\right) \frac{p_0^2+m_1^2-m_2^2}{E_1}
\nonumber \\ && \hspace{4.4cm}
+\tanh\left(\frac{\beta E_2}{2}\right) \frac{p_0^2+m_2^2-m_1^2}{E_2}
\Bigg] \nonumber \\ && \hspace{2cm}
-i\frac{\pi q_0}{2} \Bigg[
\tanh\left(\frac{\beta E_{10}}{2}\right) \frac{q_0^2+m_1^2-m_2^2}{E_{10}}
\nonumber \\ && \hspace{3cm}
+\tanh\left(\frac{\beta E_{20}}{2}\right) \frac{q_0^2+m_2^2-m_1^2}{E_{20}}
\Bigg] \Theta(q_0^2)\Bigg\}
\label{e:b2}
\end{eqnarray}
where we have abbreviated
\begin{eqnarray}
q_0^2 &=& \frac{[p_0^2-(m_1+m_2)^2][p_0^2-(m_1-m_2)^2]}{4p_0^2} \\
E_{j0} &=& \sqrt{q_0^2+m_j^2} \quad .
\end{eqnarray}
For $N^{12}(p)$ we find
\begin{eqnarray}
N^{12}(p) &=& \frac{i}{32\pi^2} \frac{1}{p_0^4} \int_0^\Lambda dq
\frac{q^2}{\left(q_0^2-q^2\right)^2}
\nonumber \\ & & \hspace{2cm} \times \left[
\tanh\left(\frac{\beta E_1}{2}\right)
\frac{(\xi + 2E_1^2)(\xi^2+4E_1^2p_0^2)-16\xi E_1^2p_0^2}{4E_1^3}
\right. \nonumber \\ & & \hspace{2.3cm}
-\tanh\left(\frac{\beta E_2}{2}\right)
\frac{\left(p_0^2-m_1^2+m_2^2\right)^2+4p_0^2E_2^2}{2E_2}
\nonumber \\ & & \hspace{2.3cm} \left.
-\frac{\beta}{\cosh^2\left(\frac{\beta E_1}{2}\right)}
\frac{p_0^2\left(q_0^2-q^2\right)\left(p_0^2+m_1^2-m_2^2\right)}{2E_1^2}
\right]
\label{e:b3}
\end{eqnarray}
with $\xi = p_0^2+m_1^2-m_2^2$.
For the function $P^{12}(k,p)$ one has
\begin{equation}
P^{12}(p,k) = -\frac{i}{4\pi^2} \int_0^\Lambda dq
\frac{q^2}{\delta_p\delta_k\delta_u} \left[
\tanh\left(\frac{\beta E_1}{2}\right)
\frac{{\cal A}\delta_u+{\cal B}\delta_k}{E_1}
+\tanh\left(\frac{\beta E_2}{2}\right)
\frac{{\cal C}\delta_p}{E_2} \right]
\label{e:b4}
\end{equation}
with
\begin{eqnarray}
\delta_p &=& p_0^2\left(p_0^2-4E_1^2\right)
\\
\delta_k &=& \left[\left(k_0^2 + m_1^2-m_2^2\right)^2-4E_1^2k_0^2\right]
          =  \left[\left(k_0^2 - m_1^2+m_2^2\right)^2-4E_2^2k_0^2\right]
\\
\delta_u &=& \left[\left(u + m_1^2-m_2^2\right)^2-4E_1^2u\right]
          =  \left[\left(u - m_1^2+m_2^2\right)^2-4E_2^2u\right]
\\
{\cal A} &=& \left[p_0^2\left(k_0^2+m_1^2-m_2^2\right)
                      + 2E_1^2(p_0^2+k_0^2-u)\right]
\\
{\cal B} &=& \left[p_0^2\left(u+m_1^2-m_2^2\right)
                      + 2E_1^2(p_0^2-k_0^2+u)\right]
\\
{\cal C} &=& \left[\left(k_0^2-m_1^2+m_2^2\right)\left(u - m_1^2+m_2^2\right)
                      - 2E_2^2(p_0^2-k_0^2-u)\right] \quad .
\end{eqnarray}
For the function $Q^{12}(k,p)$, one has
\begin{eqnarray}
Q^{12}(p,k) &=& \frac{i}{4\pi^2}\int_0^\Lambda dq \, q^2 \Bigg\{
\tanh\left(\frac{\beta E_1}{2}\right) \frac{\cal D}{2E_1^3\delta_p^2\delta_k^2}
-\tanh\left(\frac{\beta E_1}{2}\right) \frac{\cal E}{E_1\delta_p^2\delta_u}
\nonumber \\ \hspace{2cm} & &
-\tanh\left(\frac{\beta E_2}{2}\right) \frac{\cal F}{E_2\delta_k^2\delta_u}
-\frac{\beta}{4\cosh^2\left(\frac{\beta E_1}{2}\right)}
 \frac{\cal G}{E_1^2\delta_p\delta_k} \Bigg\}
\label{e:b5}
\end{eqnarray}
where
\begin{eqnarray}
{\cal D} &=& p_0^2\left[\eta \left(2E_1^2+p_0^2\right)
                        + 2E_1^2\left(4p_0^2+3k_0^2-3u\right)\right]
\nonumber \\ & & \times
             \left[\left(p_0^2+4E_1^2\right)\left(\eta^2+4k_0^2E_1^2\right)
                        + 8 \eta E_1^2 \left(p_0^2+k_0^2-u\right)\right]
\nonumber \\
         &-& 4E_1^2\left[2p_0^2\eta+\left(p_0^2+k_0^2-u\right)
                  \left(E_1^2+p_0^2\right)
             + 2p_0^2E_1^2\right]
\nonumber \\ & & \times
            \left[2p_0^2\left(\eta^2+4k_0^2E_1^2\right)
                  + \eta\left(p_0^2+k_0^2-u\right)
             \left(p_0^2+4E_1^2\right)\right]
\\
{\cal E} &=& p_0^2\left[\left(p_0^2+4E_1^2\right)\left(u+m_1^2-m_2^2\right)
          + 4 E_1^2 \left(p_0^2-k_0^2+u\right)\right]
\\
{\cal F} &=& \left(k_0^2-m_1^2+m_2^2\right)^2\left(u-m_1^2+m_2^2\right)
          + 4 k_0^2 E_2^2\left(u-m_1^2+m_2^2\right)
\nonumber \\
         &-& 4E_2^2\left(k_0^2-m_1^2+m_2^2\right)\left(p_0^2-k_0^2-u\right)
\\
{\cal G} &=& 2E_1^2\left(p_0^2+k_0^2-u\right)
          + p_0^2\left(k_0^2+m_1^2-m_2^2\right)
\label{e:b6}
\end{eqnarray}
and
\begin{equation}
\eta = k_0^2+m_1^2-m_2^2 \quad .
\label{e:b7}
\end{equation}
These expressions simplify somewhat in the case $m_1=m_2$.

\clearpage
%
%
\begin{table}
\caption[]{Parameter sets.  The parameter sets denoted I and II correspond 
to those from Ref.~\cite{bv} and which are similarly denoted therein.   PR
and
HK designate the O(3) parameter sets used in Ref.~\cite{pr} and Ref.~\cite
{hatkun} respectively.}
\label{table1}
\begin{tabular}{c|c||c|c|c|c|c}
 & Renormalization & $m_{0q}$ (MeV) & $m_{0s}$ (MeV) & $\Lambda$ (MeV) &
$G\Lambda^2$ & $K\Lambda^5$ \\ \hline
I  & Pauli-Villars & 6.2 & 175   & 795   & 2.350 & 27.83 \\
II & Pauli-Villars & 7.8 & 175   & 700   & 2.739 & 43.28 \\
PR & $O(3)$        & 5.5 & 140.7 & 602.3 & 1.835 & 12.36 \\
HK & $O(3)$        & 5.5 & 135.7 & 631.4 & 1.833 & 9.288
\end{tabular}
\end{table}

%
%
\begin{table}
\caption[]{Quark and pseudoscalar meson masses at $T=0$.  The parameter set
labels I, II, PR and HK are as described in Table~\ref{table1}. 
Experimental values are given in the final row. }
\label{table2}
\begin{tabular}{c||c|c||c|c|c|c}
   & $m_q$ (MeV) & $m_s$ (MeV) & $m_\pi$ (MeV) & $m_K$ (MeV) &
$m_\eta$ (MeV) & $m_{\eta'}$ (MeV) \\ \hline
I  & 253.0 & 489.7 & 140.8 & 521.9 & 467.6 & 845.7 \\
II & 367.8 & 557.4 & 140.0 & 485.9 & 497.3 & 1013.4 \\
PR & 367.7 & 549.5 & 135.0 & 497.6 & 514.8 & 957.7 \\
HK & 334.7 & 527.4 & 138.0 & 495.7 & 486.7 & 857.4 \\
Exp & - &    - &   138     & 495   & 548   & 958 
\end{tabular}
\end{table}

%
%
\begin{table}
\caption[]{Scalar meson masses at $T=0$. The parameter set labels I, II, PR
and HK are as described in Table~\ref{table1}.   A tentative assignment of
these
masses to experimentally measured particles is given in the final row.}
\label{table3}
\begin{tabular}{c||c|c|c|c}
   & $m_{\sigma_\pi}$ (MeV) & $m_{\sigma_K}$ (MeV) &
$m_\sigma$ (MeV) & $m_{\sigma'}$ (MeV) \\
\hline
I  & 688.8 & 904.6  & 510.7 & 1112.2 \\
II & 890.8 & 1073.4 & 735.4 & 1226.3 \\
PR & 880.2 & 1050.5 & 728.9 & 1198.3 \\
HK & 792.4 & 980.0  & 667.8 & 1149.9 \\
Exp & $a_0$(980) & $K_0^*$(1430) & $f_0$(980) & $f_0$(1300)
\end{tabular}
\end{table}

%
%
\begin{table}
\caption[]{Contributions to the scattering length $a_0^{3/2}$ in units of
           $m_\pi$ at $T=0$. The parameter set labels I, II, PR
and HK are as described in Table~\ref{table1}. }
\label{table4}
\begin{tabular}{c||c|c|c|c}
Parameter Set &  box & $t$ channel & $u$ channel & sum \\
\hline
I   &  $-0.3689$ & $0.2224$ & $0.1091$ & $-0.0375$ \\
II  &  $-0.6454$ & $0.4577$ & $0.1587$ & $-0.0289$ \\
PR  &  $-0.7047$ & $0.5332$ & $0.1741$ & $-0.0025$ \\
HK  &  $-0.6000$ & $0.4000$ & $0.1771$ & $-0.0230$ \\
\end{tabular}
\end{table}

%
%
\begin{table}
\caption[]{Contributions to the scattering length $a_0^{1/2}$ in units of
           $m_\pi$ at $T=0$. The parameter set labels I, II, PR
and HK are as described in Table~\ref{table1}. }
\label{table5}
\begin{tabular}{c||c|c|c|c}
Parameter Set & box & $t$ channel & $u$ channel & sum \\
\hline
I   &  $-0.2255$ & $0.2224$ & $0.1296$ & $0.1265$ \\
II  &  $-0.4706$ & $0.4577$ & $0.1447$ & $0.1318$ \\
PR  &  $-0.5108$ & $0.5332$ & $0.1572$ & $0.1795$ \\
HK  &  $-0.4195$ & $0.4000$ & $0.1710$ & $0.1515$ \\
\end{tabular}
\end{table}

\clearpage
%
%

%
%

%
%

%
%

%
%

%
%

%
%

%
%

%
%

%
%

\begin{references}
\bibitem{bkm1}
V. Bernard, N. Kaiser and U.-G. Mei\ss ner, Nucl. Phys. {\bf B357}, 
129 (1991).
\bibitem{bkm2}
V. Bernard, N. Kaiser and U.-G. Mei\ss ner, Nucl. Phys. {\bf B364},
283 (1991).
\bibitem{bkm3}
V. Bernard, N. Kaiser and U.-G. Mei\ss ner, Phys. Rev. {\bf D43}, 
R2757 (1991).
\bibitem{vo} M.K. Volkov and A.A. Osipov, Yad. Fiz. {\bf 34}, 1559 
(1981), Preprint JINR-P2-83-490-mc.
\bibitem{mat}
M.J. Matison {\it et al.}, Phys. Rev. {\bf D9}, 1872 (1974).
\bibitem{joh}
N.O. Johannesson and J.L. Petersen, Nucl. Phys. {\bf B68}, 397 (1973).
\bibitem{kar}
A. Karabouraris and G. Shaw, J. Phys. {\bf G6}, 583 (1980).
\bibitem{reviews} For reviews, see  U. Vogl and W. Weise,          
Prog. Part. Nucl.      
Phys.  {\bf 27}, 195 (1991);  S.P. Klevansky,  Rev. Mod. Phys.
 {\bf 64}, 649 (1992).
\bibitem{hatkun}
  T. Hatsuda and T. Kunihiro, Phys. Rep. {\bf 247}, 241 (1994).        
\bibitem{quack} E. Quack and S.P. Klevansky, Phys. Rev. {\bf C
49}, 3283
(1994).
\bibitem{dmitra}
V.~Dmitra\v{s}inovi\'c, H.J.~Schulze, R.~Tegen and R.H.~Lemmer,
Ann. Phys.~(N.Y.) {\bf 238}, 332 (1995).
\bibitem{laer1} E. Laermann, Nucl. Phys. {\bf A610}, 1c (1996).
\bibitem{laer2} E. Laermann, Nucl. Phys. {\bf B} (Proc. Suppl.)
{\bf 60A} (1998) 180.
\bibitem{muller}
J. M\"uller and S.P. Klevansky, Phys. Rev. {\bf C50}, 410 (1994).
\bibitem{lemmer}
S.P. Klevansky and R.H. Lemmer, Heidelberg Preprint HD-TVP-97/05.
\bibitem{hippe} H.-J. Hippe and S.P. Klevansky, Phys. Rev. {\bf C52}
(1995) 2172.
\bibitem{bandco} B. Bernard, U.-G. Meissner, A.H. Blin and B. Hiller, Phys.
Lett. {\bf B253} (1991) 443.
\bibitem{schulze}
H.-J. Schulze, J. Phys. {\bf G21} (1995) 185.
\bibitem{bv}
V. Bernard and D. Vautherin, Phys. Rev. {\bf D40}, 1615 (1989).
\bibitem{pr}
P. Rehberg, S.P. Klevansky and J. H\"ufner, Phys. Rev. {\bf C53}, 410 
(1996).
\bibitem{qufamily} E. Quack, P. Zhuang, Y. Kalinovsky, S.P. Klevansky
 and
J. H\"ufner, Phys. Lett. {\bf B348}, 1 (1995).
\bibitem{pdg}
R.M Barnett et al., Phys. Rev. {\bf D54}, 1 (1996).
\bibitem{petersen}
J.L.~Petersen, Phys. Rep. {\bf 2}, 155 (1971).
\bibitem{nak}
N. Nakanishi, {\it Graph Theory and Feynman Integrals\/} (Gordon and Breach,
1971).
\end{references}
\end{document}